\documentclass[fleqn,usenatbib]{mnras}

\usepackage{newtxtext,newtxmath}

\usepackage[T1]{fontenc}

\DeclareRobustCommand{\VAN}[3]{#2}
\let\VANthebibliography\thebibliography
\def\thebibliography{\DeclareRobustCommand{\VAN}[3]{##3}\VANthebibliography}

\usepackage{graphicx}	
\usepackage{amsmath}	
\usepackage{hyperref}

\title[Reflare in MAXI J1348-630]{Evolution of disc and corona in MAXI J1348-630 during the 2019 reflare: \textit{NICER} and \textit{Insight}-HXMT view}

\author[Dai et al.]{
Xiaohang Dai,$^{1}$\thanks{dai@astro.uni-tuebingen.de}
Lingda Kong,$^{1,2}$\thanks{lingda.kong@mnf.uni-tuebingen.de}
Qingcui Bu,$^{1}$
Andrea Santangelo,$^{1}$
Shu Zhang,$^{2}$
Long Ji,$^{3}$
\newauthor
Shuangnan Zhang,$^{2,4}$
Emre Seyit Yorgancioglu$^{1}$
\\
$^{1}$Institut f{\"u}r Astronomie und Astrophysik, Kepler Center for Astro and Particle Physics, Eberhard Karls, Universit{\"a}t, Sand 1, D-72076 T{\"u}bingen, Germany\\
$^{2}$Key Laboratory for Particle Astrophysics, Institute of High Energy Physics, Chinese Academy of Sciences, 19B Yuquan Road, Beijing 100049, China\\
$^{3}$School of Physics and Astronomy, Sun Yat-Sen University, Zhuhai, 519082, China\\
$^{4}$University of Chinese Academy of Sciences, Chinese Academy of Sciences, Beijing 100049, China\\
}

\date{Accepted XXX. Received YYY; in original form ZZZ}

\pubyear{2022}

\begin{document}
\label{firstpage}
\pagerange{\pageref{firstpage}--\pageref{lastpage}}
\maketitle

\begin{abstract}
In this work, using \textit{NICER} and \textit{Insight}-HXMT observations, we present a study of the broadband spectral and timing evolution of the source throughout the first reflare, which occurred about 4 months after the major outburst. 
Our findings suggest that during the reflare, below a critical luminosity $L_{\rm crit}\sim2.5\times10^{36}$ (D/2.2 kpc)$^{2}$ erg s$^{-1}$, the scale of the corona shrinks in the radial direction, whereas the inner radius of the disk does not change considerably; however, the inner radius of the disk starts to move inward when the source exceeds the critical luminosity.
We conclude that at low luminosity the increase in accretion rate only heats up the inner zone of the accretion disc without the transfer of angular momentum which occurs above a certain luminosity.
\end{abstract}




\section{Introduction}

X-ray binaries (XRBs) consist of a black hole (BH) or Neutron Star (NS) gravitationally bound to a normal companion star. 
Depending on the mass of the donor stars, XRBs are classified as high-mass X-ray binaries (HMXBs: $M \gtrsim 5M_{\odot}$) and low-mass X-ray binaries (LMXBs: $M \lesssim 1M_{\odot}$). 
In HMXBs, the compact objects accrete mainly through the stellar wind from the companion stars, whereas in LMXBs, accretion is in the form of Roche-lobe overflow (\citealp{Shakura1973AA, Reig2011ApSS}) of the companion.
Four stable accretion flow structures can exist: a geometrically thin, optically thick, cold accretion disk (\citealp{Shakura1973AA}); a geometrically thick, optically thin, hot Corona (\citealp{Gilfanov2010LNP}); an Advection-Dominated Accreting Flow (ADAF)  (\citealp{Narayan1995ApJ, Abramowicz1995ApJ}); and an optically-thick advection dominant accretion flow, often called "slim disk", which is expected when the mass accretion rate is very high (\citealp{Kawaguchi2003ApJ...593...69K}).

Some black hole transients spend virtually all of their lives in a fairly faint state in the X-ray band, but they occasionally experience unpredictable and irregular active X-ray outbursts that last from a few days to several months (\citealp{Belloni2011BASI}). 
In contrast, some black hole binary systems remain active in the X-ray band for rather long periods of time, such as GRS 1915+105 (\citealp{Belloni2000AA}) and Cyg X-1 (\citealp{Liang1984SSRv}).
LM-BHXRBs in outbursts are characterized by the rapid evolution of their spectral and timing properties, which can be studied in terms of different states (\citealp{Homan2005ApSS, Remillard2006ARAA}).
A classical black hole binary outburst may cycle through many different states and follow a "q" shape in the Hardness Intensity Diagram (HID), which quickly and clearly displays the relationship between the distribution of soft and hard photons in the energy spectrum and the source intensity. 
Furthermore, in different branches of the HID, the quasi-periodic and aperiodic signals in the light curve can produce intriguing different power density spectrum (PDS) in the frequency domain.

According to these properties, LM-BHXRBs are typically classified according to different states (\citealp{Belloni2010LNP, Remillard2006ARAA, Motta2009MNRAS}). 
At the beginning of an outburst, the flux rises rapidly, the majority of emitted photons are in the hard X-ray range, and the spectra are dominated by a hard power-law component in the X-ray band (power-law photon index $\Gamma$ between $\sim$ $1.5-1.7$) in what is known as the hard state (HS). 
A power density spectrum (PDS) in the HS normally has a strong band-limited noise (30\%$\sim$40\% RMS) component accompanied by a low-frequency Quasi-Periodical Oscillation (QPO).
Following the HS, the outburst switches to the intermediate state (IMS) which can further be classified into a hard-intermediate state (HIMS) and soft-intermediate state (SIMS).
In this transition state, the emission gradually changes from power law to thermal. 
Inverse Compton photons scattering in a hot corona or the jet base, which originated from a geometrically thick, optically thin region near the compact object determines the power-law component in the energy spectrum.
The band-limited noise decreases with increasing flux, and the total RMS of HIMS (10\%$\sim$20\%) is lower than that in the HS. 
Compared to the HIMS where type-C QPOs are normally present, a PDS in the SIMS has a RMS lower than 10\% and a weak power law component with the occasional presence of type-A or type-B QPOs.
After the IMS, photons mainly originate from the disk, which displays a thermal component in the energy spectrum. Then, the transient switches into a soft state (SS), where the spectrum is the softest and dominated by thermal emission. The spectrum can be described by a strong multi-color blackbody component with a weak power-law tail, and the PDS shows a further weakened power-law noise component. 
Generally, it is assumed that the disk truncated radius gets to the innermost stable circular orbit (ISCO) during the SS.
This is consistent with observations showing that the disk component gradually gets larger during this period.
After the SS, the source once again begins to harden and return to the IMS. The IMS this time is soft-to-hard state switches. 
Finally, the outburst enters HS again, and it quenches after a few days, completing the hysteresis cycle in the HID.

The truncated accretion disk and hot corona cooling are successfully used to illustrate spectro-temporal variability, jet emission, and disk wind in the vicinity of accreting black hole transients (\citealp{Esin1997ApJ, Meyer-Hofmeister2009AA, Done2007AARv, Ponti2012MNRAS}). 
The Lense-Thirring precession model of the hot corona seems consistent with observations showing increasing QPO with frequency decreasing inner disk radius and increased QPO fractional RMS with higher energies. (\citealp{Ingram2009MNRAS}). However, some recent new observational evidence does not fit the truncated disk model. In MAXI J1820+070, \cite{Kara2019Natur} found that the broadened iron K emission line keeps constant profiles in the HS, while the timescale of the reverberation lags shortens by an order of magnitude over a period of weeks. This suggests that in the HS, there is a reduction in the spatial extent of the corona, rather than a change in the inner edge of the accretion disk. Later, in the same source, \cite{You2021NatCo} found that the Compton hump arises from disk reflection, indicating a jet-like corona outflowing faster, and \cite{Ma2021NatAs} found a QPO with large RMS and soft lag up to 200 keV suggesting that the QPO probably originates from the precession of a small-scale jet. Therefore, up to now, the nature of the disc-corona-jet is still unclear.

Sometimes, after their main outburst, the sources' flux shows several re-brightening episodes, reaching a level of a few orders of magnitude lower than the peak flux of the main outburst. This phenomena is called a ``reflare'' or ``mini-outburst'' (\citealp{Patruno2016ApJ, Cueneo2020MNRAS, Zhang2020MNRAS}).
In these reflares, the flux is at least approximately one to two orders of magnitude fainter than that of the main outburst, but the whole process has a similar behavior in HS and IMS. 
In GRS 1739–278 and MAXI J1535-571, \cite{Yan2017MNRAS} and \cite{Cueneo2020MNRAS} found that some reflares also showed hysteresis loops in the hardness-intensity diagram (HID). 
 \cite{Yan2017MNRAS} found that the luminosity in the hard-to-soft and HS of the mini-outbursts also follow the correlation that can be previously found in main outbursts.
The similarities between the main outbursts and followed ``nesting dolls-like'' reflares point out the same physical origin for both phenomena. But, how the ignition of reflares occur after the main outburst runs out of  accumulated matter at the outer accretion disc is still an unsolved question. 

MAXI J1348-630 is a black hole candidate that was discovered with the Monitor of an All-sky X-ray Image Gas Slit Camera (MAXI/GSC) at the start of its outburst on January 26, 2019 (\citealp{Yatabe2019ATel, Negoro2020ATel, Tominaga2020ApJ}), and was subsequently observed by the Neil Gehrels Swift Observatory (Swift) (\citealp{DElia2019GCN1}), INTEGRAL (\citealp{Lepingwell2019ATel, Cangemi2019ATel}), and the Neutron star Interior Composition Explorer (\textit{NICER}) (\citealp{Sanna2019ATel}). \textit{Insight}-HXMT (\citealp{Chen2019ATel}) observed both the main outburst and mini outburst in the X-ray bands.
Such outburst encouraged multi-wavelength observations, i.e., the 50 cm T31 telescope (\citealp{Denisenko2019ATel}), the Las Cumbres Observatory 2 m and 1 m robotic telescopes, and the Southern African Large Telescope in the optical bands; the Australia Telescope Compact Array (ATCA) (\citealp{Russell2019ATel}), MeerKAT, and the Murchison Widefield Array in the radio bands.
A distance of $\sim 2.2$ kpc was estimated from the H I absorption (\citealp{Chauhan2021MNRAS}) or of $\sim 3.39$ kpc from the dust-scattering ring around the source which was calculated by data combination from SRG/eROSITA, \textit{XMM-Newton}, MAXI, and \textit{Gaia}(\citealp{Lamer2021AA}). Following its main outburst, the system underwent several reflares before returning to the quiescent state.

In this paper, we illustrate and discuss the broadband spectral evolution and timing properties of the first reflare of MAXI J1348-630 immediately after its main outburst. With the large effective area of \textit{NICER} and \textit{Insight}-HXMT at low and high energies, respectively, and their continuous coverage of the burst, we can study the spectral and timing properties of the disk and corona during the reflare.
We report the observations and the data reduction in Section 2, the reflare evolution results in Section 3, and discuss the implication of the results in Section 4. The summary and conclusions are presented in Section 5.

\section{Observations and Data reduction}

\subsection{\textit{Insight}-HXMT}

The Hard X-ray Modulation Telescope (\citealp{Zhang2014SPIE, Zhang2020SCPMA}), named \textit{Insight}-HXMT, was launched on June 15, 2017. 
It has three primary payloads with high temporal resolutions and sizable effective areas: the Low Energy X-ray Telescope (LE: 384 $\rm cm^2$ at 1-10 keV, \citealp{2020SCPMA..63x9505C}); the Medium Energy X-ray Telescope (ME: 952 $\rm cm^2$ at 5-30 keV, \citealp{2020SCPMA..63x9504C}); the High Energy X-ray Telescope (HE: 5000 $\rm cm^2$/20-250 keV, \citealp{2020SCPMA..63x9503L}).
The Fields of View (FoVs) are $1.6^{\circ}\times6^{\circ}$, $1^{\circ}\times4^{\circ}$, and $1.1^{\circ}\times5.7^{\circ}$ for LE, ME, and HE, respectively.
Due to its broadband capacity ($1-250$ keV) and large collection area at high energies, \textit{Insight}-HXMT is well suited to study the timing and spectral properties of non-thermal emission from X-ray sources.

\textit{Insight}-HXMT was triggered to observe MAXI J1348-630 from January 27, 2019 (MJD 58510) to July 29, 2019 (MJD 58693), starting one day after its discovery and included both its first and the second outbursts (or first reflare). The main outburst lasted for 108 days from MJD 58509 to MJD 58617 (a count rate of $2$ ct s$^{-1}$).
Then, the second outburst lasted for 56 days, between MJD 58637 to MJD 58693.
We obtained 32 observations from \textit{Insight}-HXMT (a total of 73 exposures), with 20 observations having more than one exposure. 

We screen the event data and extract the light curves and spectra with the \textit{Insight}-HXMT Data Analysis Software (HXMTDAS) v2.04 and the calibration model v2.05\footnote{http://hxmtweb.ihep.ac.cn/software.jhtml}.
For observations with more than one exposure in one day, we combine their spectra to one spectrum by ftools \textit{addspec}, \textit{addrmf}, and \textit{addarf}. 
The energy bands considered for spectral analysis are $2-10$keV for LE, $10-30$keV for ME, and $30-200$keV for HE. The background spectra and response files are created with LEBKGMAP, MEBKGMAP, and HEBKGMAP tasks and LERSPGEN, MERSPGEN, and HERSPGEN, respectively, based on the latest standard \emph{Insight}-HXMT background models and calibration (\citealp{Liao2020a, Guo2020JHEAp, Liao2020b}).

\subsection{\textit{NICER}}

The X-ray Timing Instrument (XTI) of the Neutron Star Interior Composition Explorer (\textit{NICER}) was launched aboard a SpaceX Falcon 9 rocket on June 3, 2017. 
It comprises an aligned collection of 56 X-ray ''concentrator'' optics (XRC) and silicon drift detector (SDD) pairs.
It provides high signal-to-noise-ratio photon-counting capability within the $0.2-12$keV X-ray band with a peak effective area of $\sim$ 1900 cm$^{2}$ at 1.5 keV, which is perfectly matched to black body component in low temperatures, such as surface radiation from the neutron star or accretion disk radiation from the black hole.

\textit{NICER} started to observe MAXI J1348-630 after its discovery several hours later. 
We use \textit{NICER} data between MJD 58637 to MJD 58693, which are on nearly identical days of \textit{Insight}-HXMT observations.
We apply \textit{NICER} Level2 standard calibration and filtering to process data by \textit{\textit{NICERl2}}, and we get the spectral response files by \textit{NICERRMF} and \textit{NICERARF}. 
We use the ``nibackgen3C50'' model to estimate the background for the spectral analysis. We remove data from detectors 14 and 34 to escape occasional episodes of increased electronic noise.

Other analyses are processed using HEASOFT version 6.29. We extract light curves using XSELECT with 0.001s bin size in $1-10$ keV energy band. 
We also extract the light curve in $2-4$ keV and $4-7$ keV to calculate the hardness ratio.

\subsection{Spectral and timing analysis}

We jointly fit \textit{NICER} spectra, with ME  HE of \textit{Insight}-HXMT spectra for each data segment for spectral analysis. 
We produce an average power density spectra (PDS) for timing analysis in the $10-30$keV and $25-80$keV energy band for each data segment from HXMT. 
Time resolution and time intervals are 0.004s and 100s, respectively, corresponding to the frequency range $0.01-125$ Hz.
The Miyamoto normalization is used in PDS, and we use Xspec v12.12.0 (\citealp{Arnaud1996}) to fit PDS and estimate the root mean square (RMS) with a sum of Lorentzian components for red noise and QPO. An additional power-law component, with zero index, is used to model the white noise. 
We use the cross-correlation method in the Stingray\footnote{https://docs.stingray.science} package to get the time lag in different frequency and energy bands (\citealp{matteo_bachetti_2022_6290078, Huppenkothen2019, 2019ApJ...881...39H}). 

For spectral fitting in Xspec, the parameter uncertainties are reported at the 90\% confidence interval.
Considering the current accuracy of the instrument calibration and appropriate $\chi^2$, we add the 3\% system error for \textit{NICER} $0.3-1.5$ keV, 0.5\% for \textit{NICER} $1.5-10$ keV, 1\% for ME $8-40$ keV, and 1\% for HE $30-150$ keV during spectral fittings. 
We use \textit{ftgrouppha} from ftools based on \cite{Kaastra2016AA} with the signal-to-noise $\geq 5$ to group and improve the spectrum quality for \textit{NICER} (grouptype=optsnmin groupscale=5.0) and \textit{Insight}-HXMT (grouptype=snmin groupscale=5.0).

\section{Results}

\subsection{Lightcurve of the reflare}

\begin{figure*}
    \centering\includegraphics[scale=0.8]{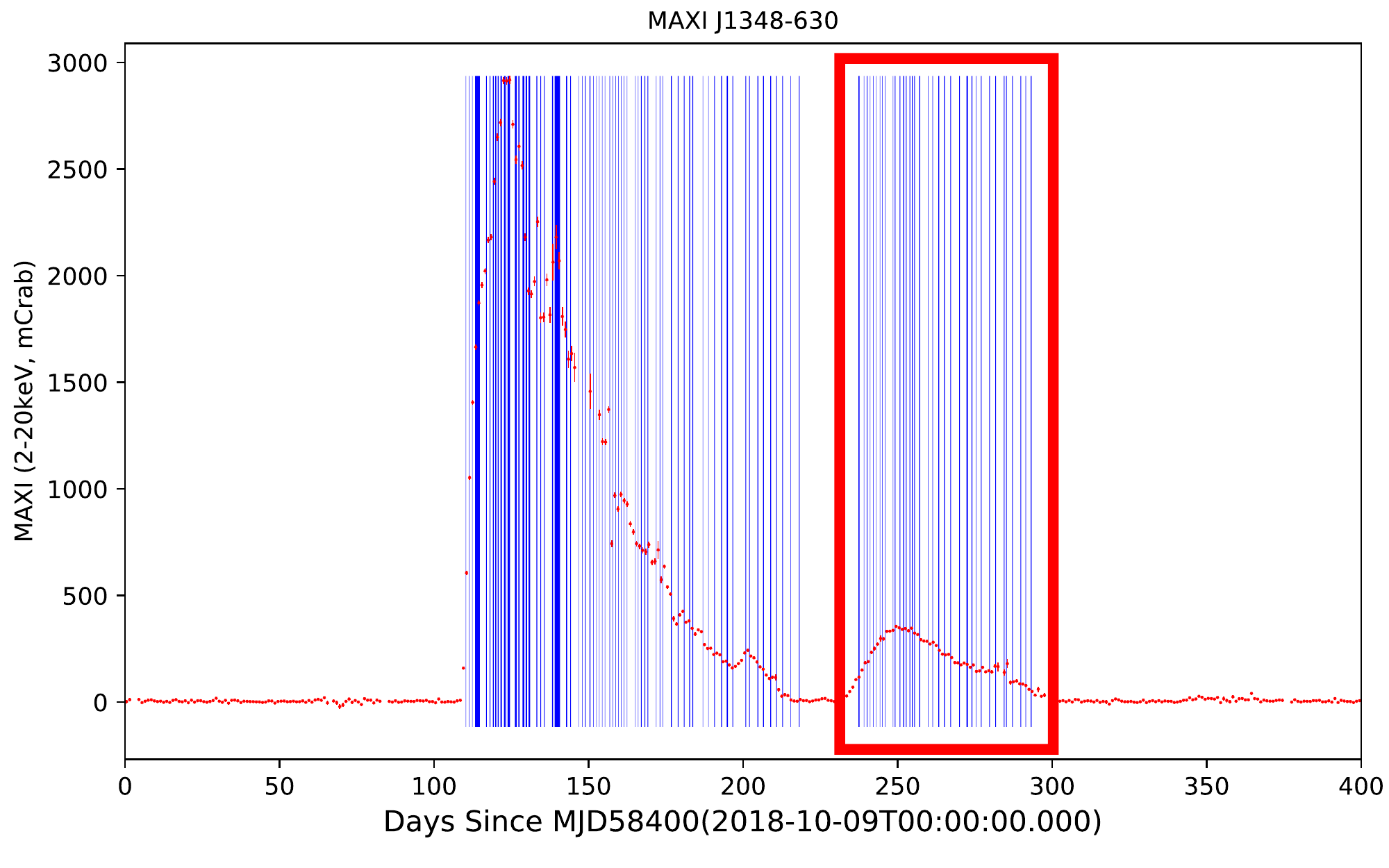}
    \caption{The MAXI $2-20$keV light curve shows the main outburst and the first reflare of MAXI J1348-630. The blue areas indicate the \textit{Insight}-HXMT's observations. The red rectangle displays the focused reflare part.}
    \label{Fig1. maxicurve}
\end{figure*}

\begin{figure*}
    \centering\includegraphics[scale=0.75]{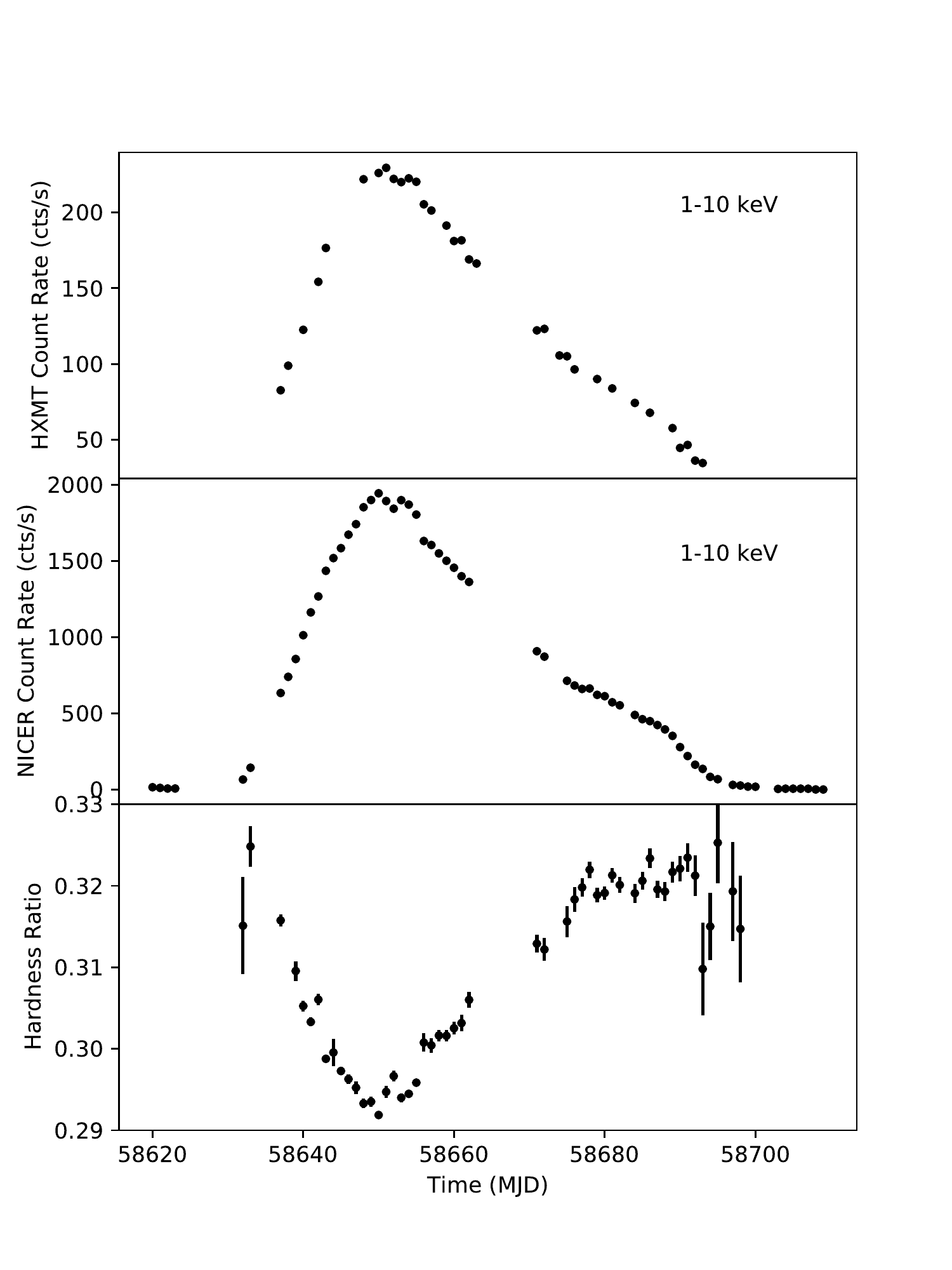}
    \caption{Upper: HXMT $1-10$keV light curve. Medium: \textit{NICER} $1-10$keV light curve. Bottom: The hardness ratio of ($4-7$)keV/($2-4$keV) from \textit{NICER}'s observations.}
    \label{Fig2. detailcurve}
\end{figure*}

Figure~\ref{Fig1. maxicurve} shows the light curve of both the main outburst and second outburst from the MAXI ($2-20$keV) all-sky monitoring of MAXI J1348-630.
After the main outburst, the source remained in quiescence for 20 days, which is less than one-fifth of the main outburst duration and nearly one-third of the second outburst duration.
The second outburst lasted for 56 days from MJD 58637 to MJD 58693 and was one order of magnitude fainter than the peak luminosity of the main outburst. 
Hence, it has all the features of a "reflare".
Figure~\ref{Fig2. detailcurve} shows the light of the reflare in detail, and the trends of $1-10$ keV light curves from \textit{Insight}-HXMT/LE and \textit{NICER} are consistent. 
The flux of the reflare rapidly rises and slowly decays like other black hole sources. 
To confirm its characteristics, we use \textit{NICER} observations to define the hardness ratio (HR) from the $4-7$keV to $2-4$keV count rates, since \textit{NICER} performed much more observations than \textit{Insight}-HXMT. 
The HR during the reflare reaches lower values. In general evolve with the reflare and ranges in the window $\sim\ 0.29-0.33$, as shown in the bottom panel in Figure~\ref{Fig2. detailcurve}. 
The HR shows an anti-correlation with the flux, which means a softening process of spectra during the outburst.  

\subsection{Spectral analysis}

\begin{figure*}
    \centering\includegraphics[scale=0.8]{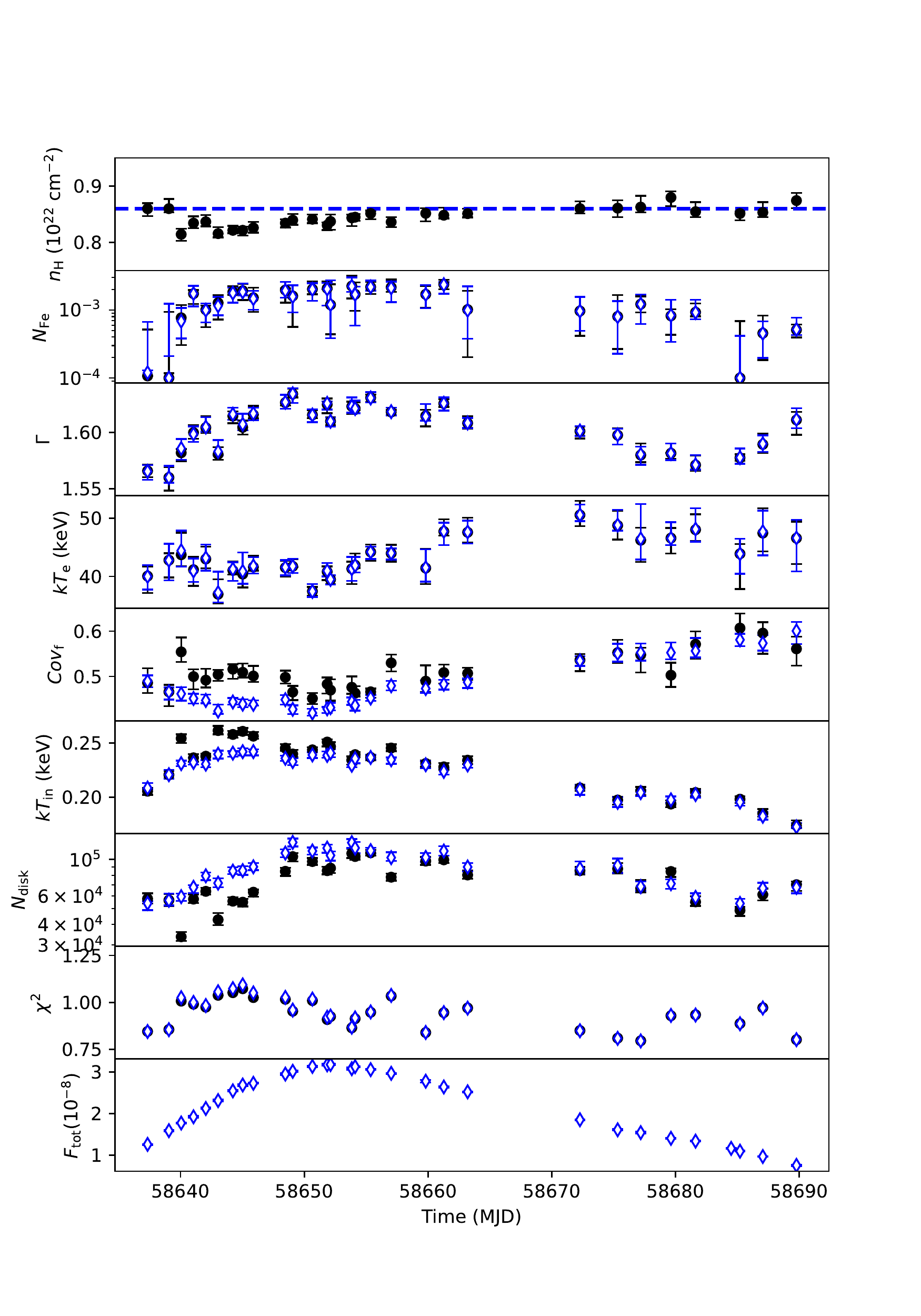}
    \caption{Spectral parameter evolution with time. 
    The panel shows the fitting result of \textit{TBabs$\times$(gaussian+gaussian+thcomp$\times$diskbb)}, and the black and blue points are on behalf of the free $n_{\rm H}$ and fixed $n_{\rm H}$, respectively. 
    $F_{\rm tot}$ is the total unabsorption flux in $0.1-150$ keV.}
    \label{Fig3. Parameters_diskbb}
\end{figure*}

When performing the spectral analysis, we combined \textit{NICER} and \textit{Insight}-HXMT observations, using \textit{NICER} observations between $0.3-10$ keV to better constrain the column density, and \textit{Insight}-HXMT observations between $8-150$ keV to constrain the physical characteristics of non-thermal emission throughout the reflare epoch (MJD 58637- MJD 58693).
Because of the narrow allowed energy ($2-7$ keV) band and lower effective area than \textit{NICER}, and for some observations (on MJD 58640, 58642, 58645, and 58646), due to the lack of GTI, \textit{Insight}-HXMT cannot provide the low energy band below 10 keV, though this is compensated with \textit{NICER}'s observations. 
Therefore, we only use \textit{NICER} observations for our spectral analysis in the lower energy band.
In general, the spectrum from the black hole binaries can be typically fitted by a multicolor blackbody \textit{diskbb} from the accreting disk (\citealp{Shakura1973AA}), the non-thermal component from the comptonization in a corona (\citealp{Sunyaev1980AA, Sunyaev1985AA}), which can be described by a phenomenological \textit{cutoffpl} model, or physical models such as \textit{compTT} (\citealp{Titarchuk1994ApJ}) and \textit{thcomp} (\citealp{Zdziarski1996MNRAS.283..193Z, Zdziarski2020MNRAS.492.5234Z}), and, sometimes, an additional relativistic reflection spectrum from an illuminated accretion disk (\citealp{Fabian2016AN....337..375F}).
Firstly, we fit the spectra with the model \textit{TBabs$\times$(Gaussian+Gaussian+thcomp$\times$diskbb)}, where \textit{thcomp} is a much better physical self-consistent model of the continuum shape from thermal Comptonization than an exponential cutoff power-law but has similar corresponding free parameters, such as the spectral index $\Gamma$, and the electron temperature ($kT_{\rm e}$) (\citealp{Zdziarski2020MNRAS}). 
The model also provides an accurate description of the Comptonization process at energies comparable to those of seed photons from the accretion disk; hence it can afford a reasonable disk temperature $kT_{\rm in}$ and normalization $N_{\rm bb}$. 
The Tuebingen-Boulder ISM absorption model \textit{TBabs} is used (\citealp{Wilms2000}), and the equivalent hydrogen column density $n_{\rm H}$ is first set as a freely varying parameter during the fit. 
A \textit{gaussian} line is fixed at 6.4 keV to reproduce the Fe-K line; the width of the Fe-K line has been fixed at 0.3 keV, which is the average value from fittings.
Additional residuals are observed around 1 keV, which is possibly due to instrumental effects (\citealp{Zhang2020MNRAS, Kumar2022MNRAS}). 
Hence, we take the same approach as \cite{Zhang2020MNRAS} and add another \textit{gaussian} component to account for these residuals; the centroid energy of the line was fixed at 1 keV, and the width of the line was fixed at 0.9 keV, which is the average value from fittings.
These additional \textit{gaussian} components have a significant effect on the \textit{diskbb} normalization $N_{\rm disk}$ but have little effect on the evolution of $kT_{\rm in}$ and photon index $\Gamma$. 
The black points in Figure~\ref{Fig3. Parameters_diskbb} show the results of the best fits. The column density $n_{\rm H}$ basically remains stable around $\sim\ 0.86\times10^{22}$ cm$^{-2}$, except near the peak of the outbreak. 
We use the $n_{\rm H}$ value from the average value of the spectral fitting, and it is also consistent with Swift/XRT results during the main outburst (\citealp{Tominaga2020ApJ, Carotenuto2021MNRAS_A, Carotenuto2021MNRAS_B}). 
Since the $n_{\rm H}$ can easily affect the measurement of $kT_{\rm in}$, we keep it fixed at this value in the following analysis processes; the results are shown in Figure~\ref{Fig3. Parameters_diskbb} with blue hollow points.
$N_{\rm Fe}$, $\Gamma$, $kT_{\rm in}$, and the \textit{diskbb} normalization $N_{\rm bb}$ increases with the flux, while the trends of the covering fraction $Cov_{\rm f}$ decreases with the flux.  
The multi-color blackbody component \textit{diskbb} describes emission arising from the accretion disk with the defined normalization parameter $N_{\rm disk}=(R_{\rm in}/D_{10})^{2}cos\rm \theta$, where $R_{\rm in}$ is an apparent inner disk radius. 
According to our results, the disk recedes at the peak flux, which is similar to the results in \cite{Zhang2020MNRAS}.
We observe here that although the 1 keV gaussian component can indeed affect the value of the $N_{\rm disk}$, it will not significantly change the trend.
Therefore, it would exist that a scenario in which the disk is further away from the compact object during the outburst, is not consistent with the canonical scenario.
As the flux increases, the $Cov_{\rm f}$ drops from 0.5 to 0.4, indicating that at the peak of the outburst, fewer photons scatter with the high-temperature electrons in the corona.
At the same time, $\Gamma$ increases with the flux from $\sim$ 1.55 to 1.65 consisting of the variation of the hardness ratio (HR) in HID.
At the end of the reflare, there is an insignificant upward trend of the $kT_{\rm e}$.

\begin{figure*}
    \centering\includegraphics[scale=0.8]{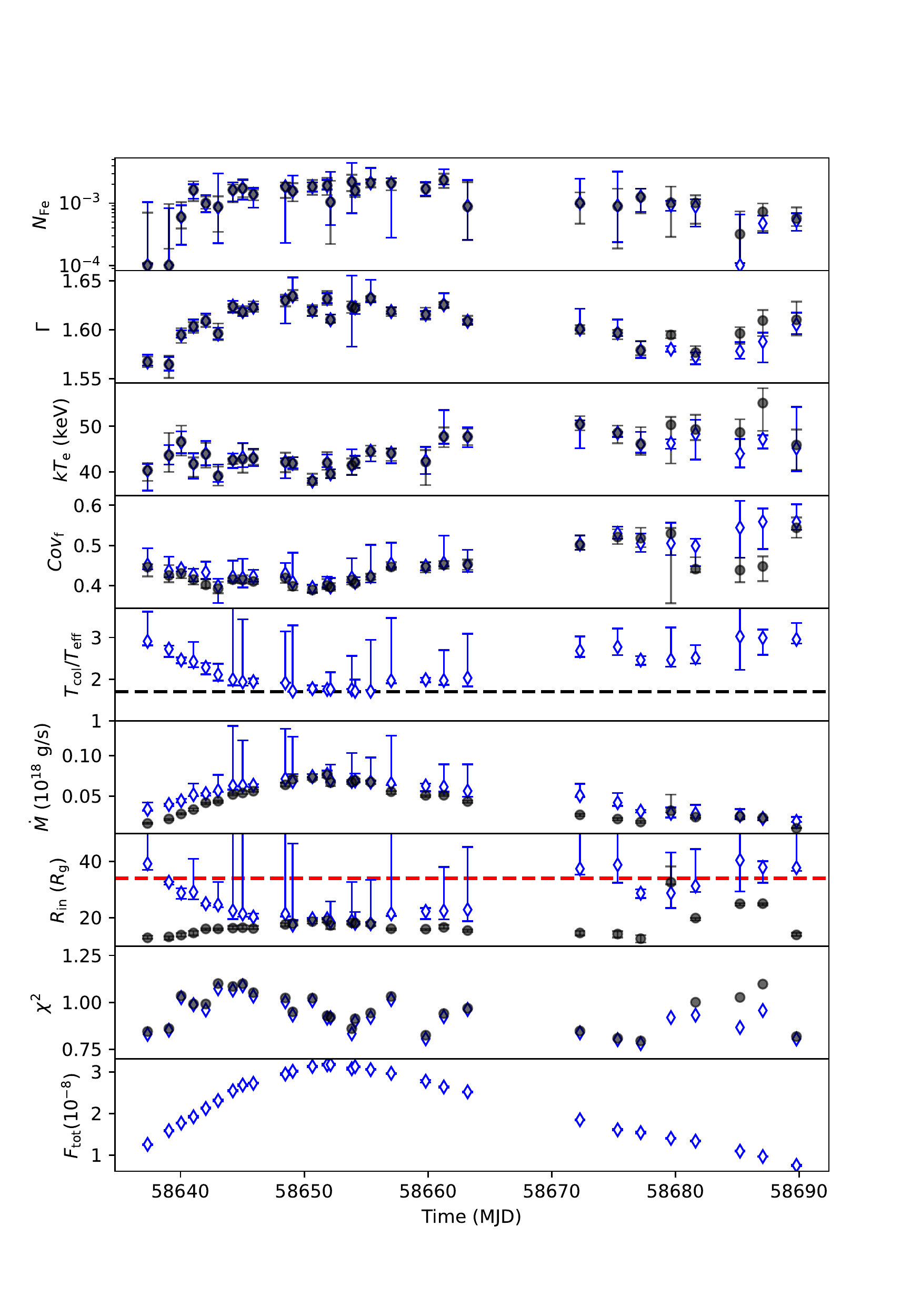}
    \caption{Spectral parameter evolution with the model \textit{TBabs$\times$(gaussian+gaussian+thcomp$\times$kerrd)} with fixed $n_{\rm H} = 0.86\times10^{22}$ cm$^{-2}$}. The blue and black points are on behalf of a free harden factor $T_{\rm col}/T_{\rm eff}$ and fixed harden factor $T_{\rm col}/T_{\rm eff}=1.7$, respectively (shown in black dashed line). The red dashed line represents the $R_{\rm in}=34$ $R_{\rm g}$. The time range of stable $R_{\rm in}$ is consistent with the time range with the disk flux below $3.3\pm0.2\times 10^{-9}$ erg cm$^{-2}$ s$^{-1}$ in Figure~\ref{Fig6. flux_T}. 
    \label{Fig4. Parameters_kerrd}
\end{figure*}

In Figure~\ref{Fig4. Parameters_kerrd}, we show the fitting results from another model \textit{TBabs$\times$(gaussian+gaussian+thcomp$\times$kerrd)}, where \textit{kerrd} describes the thermal emission from the optically thick accretion disk around a Kerr black hole (\citealp{Laor1991ApJ, Ebisawa2003ApJ}); MAXI J1348-630 is a high spin system with $a=0.80\pm0.02$ (\citealp{Kumar2022MNRAS}).
We fix the disk inclination angle at $36.5^{\circ}$, the mass of the central object at $8.7\pm0.3$ $M_{\odot}$ (\citealp{Kumar2022MNRAS}), and the distance at 2.2 kpc (\citealp{Chauhan2021MNRAS}). 
Firstly, we fix the spectral hardening factor, $T_{\rm col}/T_{\rm eff}$ at 1.7 for a typical accretion disk with high values of the accretion rate and low values of the viscosity parameter (\citealp{Shimura1995ApJ, Ebisawa2003ApJ}), where $T_{\rm eff}$ is the effective temperature of the accretion disk and the $T_{\rm col}$ is the color temperature of the accretion disk. Because of the Comptonization in the disk atmosphere, the color temperature of the local emission becomes higher than the effective temperature. And we left the inner radius $R_{\rm in}$ and mass accretion rate $\dot{M}$ as free parameters during the fittings.
The parameters are shown in Figure~\ref{Fig4. Parameters_kerrd} with black points.
$N_{\rm Fe}$, $\Gamma$, $kT_{\rm e}$, and $Cov_{\rm f}$ follow similar trends and values as in the previous model. 
$R_{\rm in}$ follows a similar trend as $N_{\rm disk}$ when the accretion rate increases, moving outward from $\sim10$ $R_{\rm g}$ to $\sim17$ $R_{\rm g}$. 
Considering that the change of $T_{\rm col}/T_{\rm eff}$ itself may affect the evolutionary trend and measurements of $R_{\rm in}$, we further set $T_{\rm col}/T_{\rm eff}$ free to change and plot the results in Figure~\ref{Fig4. Parameters_kerrd} with blue points.
We found that the values of $\dot{M}$ are higher, but the evolutionary trend is unchanged. The estimates of $T_{\rm col}/T_{\rm eff}$ and $R_{\rm in}$ are greatly different from those before. 
The $R_{\rm in}$ stays stable at $\sim34$ $R_{\rm g}$ in the low flux level from MJD 58670, while the $R_{\rm in}$ goes inward to $\sim17$ $R_{\rm g}$ near the reflare peak.
And the values of $T_{\rm col}/T_{\rm eff}$ also show a significant deviation, except around the reflare peak; $T_{\rm col}/T_{\rm eff}$ decreases from 3 to 1.7 to reach again $ \sim 3$ at the end of reflare. 

\begin{figure*}
    \centering\includegraphics[width=0.49\textwidth]{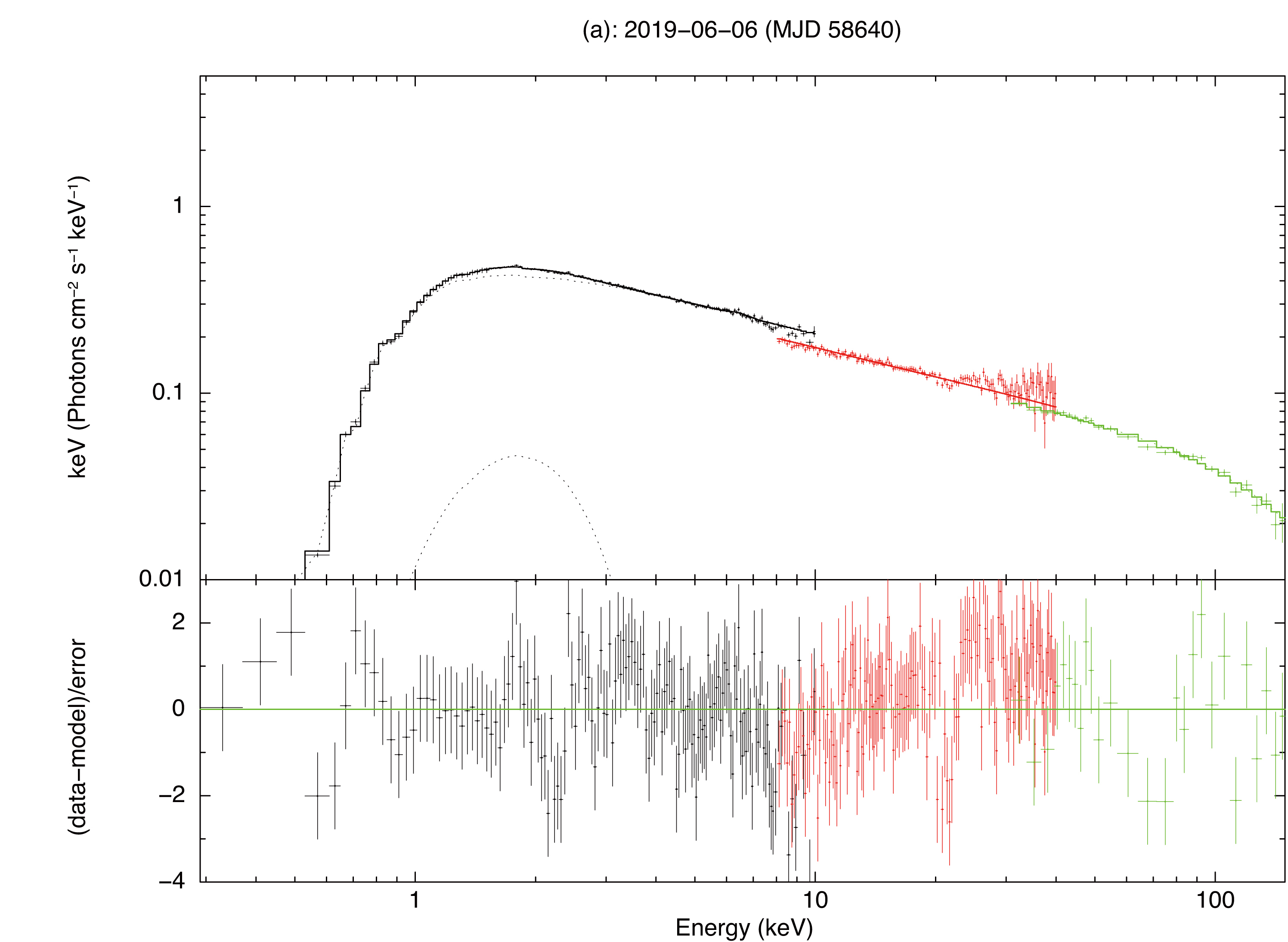}
    \centering\includegraphics[width=0.49\textwidth]{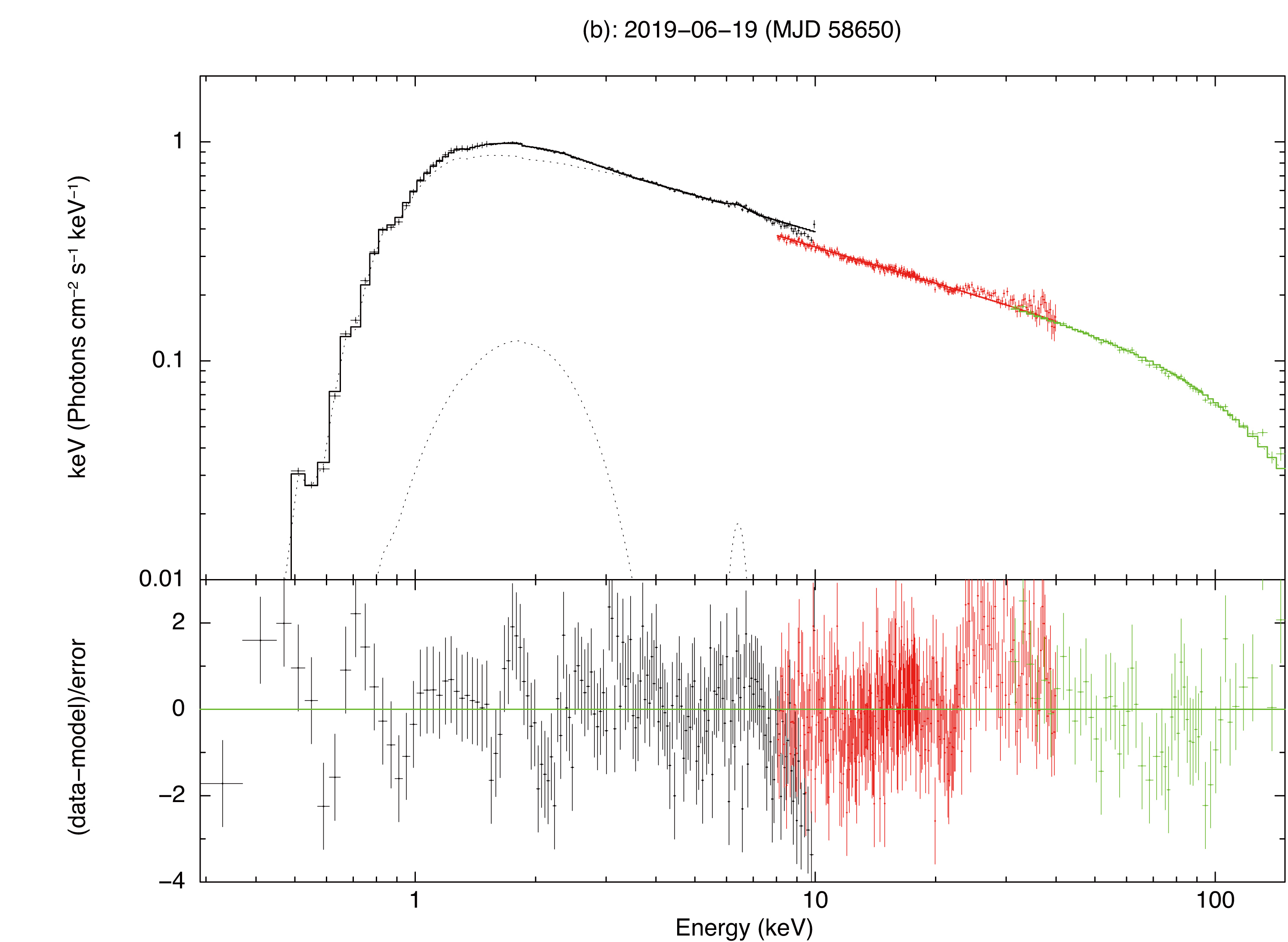}
    \caption{(a): present the spectrum on MJD 58640 with the model \textit{TBabs$\times$(gaussian+gaussian+thcomp$\times$kerrd)} with the fixed $n_{\rm H} = 0.86\times10^{22}$ cm$^{-2}$} and free harden factor respectively. (b): present the spectrum on MJD 58650 from \textit{Insight}-HXMT with the model \textit{TBabs$\times$(gaussian+gaussian+thcomp$\times$kerrd)} with the fixed $n_{\rm H} = 0.86\times10^{22}$ cm$^{-2}$ and free harden factor respectively. The black, red and green points mark the spectrum of \textit{NICER} ($0.3-10$ keV), ME ($8-40$ keV) and HE ($30-150$) keV.
    \label{Fig5. spectral fitting}
\end{figure*}

Figure~\ref{Fig5. spectral fitting}(a) and Figure~\ref{Fig5. spectral fitting}(b) shows the fitting results at the peak flux (on MJD 58650) with the model \textit{TBabs$\times$(gaussian+gaussian+thcomp$\times$diskbb)} and \textit{TBabs$\times$(gaussian+gaussian+thcomp$\times$kerrd)}, respectively.
For spectral fitting with fixed the $n_{\rm H}$, fitting errors $\chi^{2}$ are constrained in compatible values $\sim\ 1.2$. 
The additional dips $\sim\ 0.56$ keV and $\sim\ 0.87$ keV are caused by known features of \textit{NICER} around the O-K edge and the Ne-K edge. The peak at $\sim\ 1.84$ keV is due to the Si-K edge.
These systematic features are more visible only in a few observations where the count rate is relatively high and do not affect the measurement results of the continuum components.

\begin{figure*}
    \centering\includegraphics[width=0.49\textwidth]{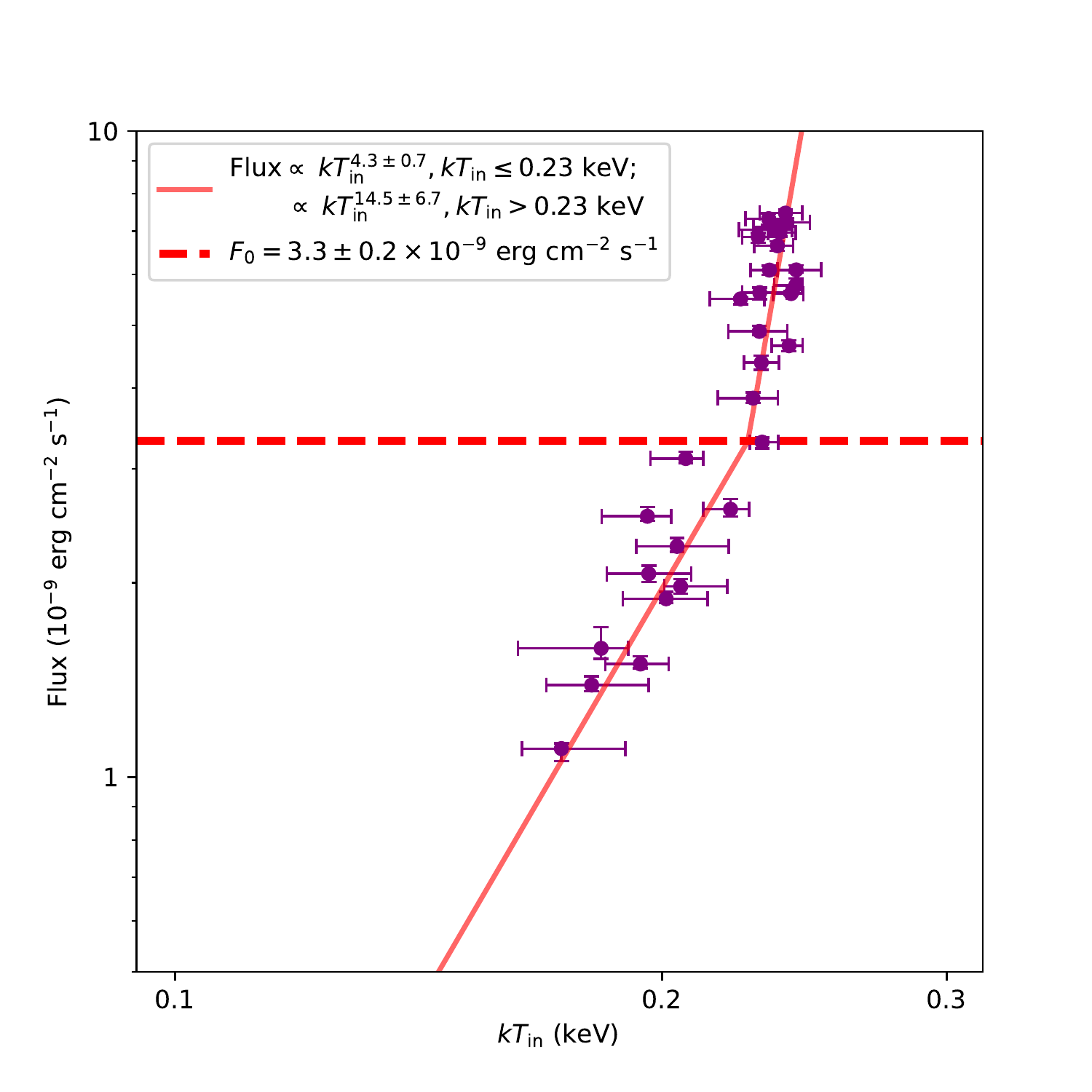}
    \centering\includegraphics[width=0.49\textwidth]{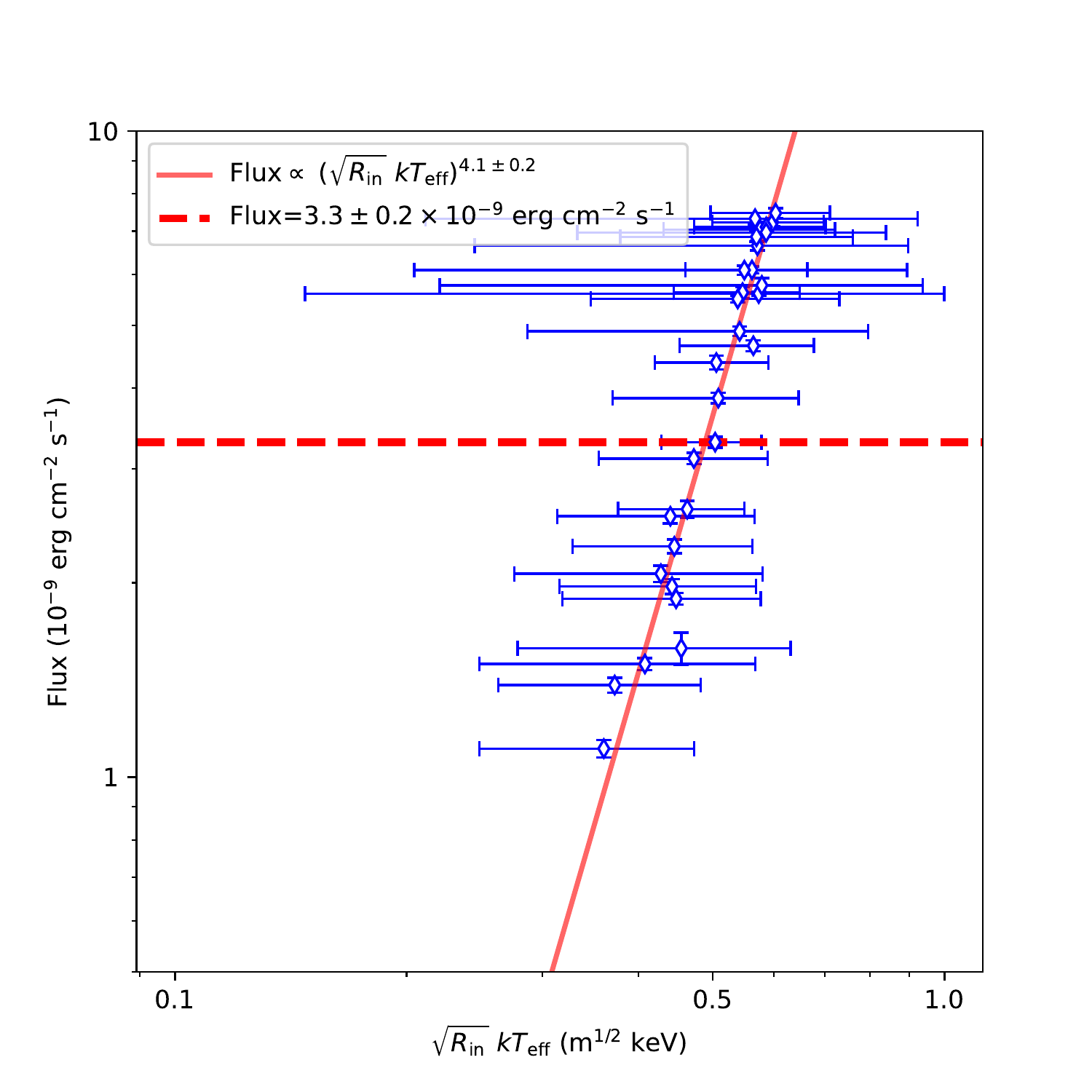}
    \caption{Left panel: Evolution of unabsorbed disk flux with disc temperature. The unabsorbed disk flux is obtained from the cflux model calculated between 0.1 keV to 150 keV. We use a broken power-law model to fit the data. And we find that the points follow the F $\propto kT_{\rm in}^{4.3\pm0.7}$ (the fluxes obey the $kT_{\rm in}^{4}$ law assuming $R_{\rm in}$ is not change) when the $kT_{\rm in}\leq0.23$ (the flux below $3.3\pm0.2 \times10^{-9}$ erg cm$^{-2}$ s$^{-1}$). While, when $kT_{\rm in} > 0.23$, the fluxes do not obey the $kT_{\rm in}^{4}$ law.}
    Right panel: Evolution of unabsorbed disk flux with $\sqrt{R_{\rm in}}\ kT_{\rm eff}$. The $kT_{\rm eff}$ is $kT_{\rm in}/f_{\rm color}$, where the hardening factor $f_{\rm color}=T_{\rm col}/T_{\rm eff}$ and the inner radius $R_{\rm in}$ are get from the \textit{kerrd} model. After taking into account changes in $R_{\rm in}$ and hardening factor, all points complied with the F $\propto R_{\rm in}^2\ T_{\rm eff}^{4}$.
    \label{Fig6. flux_T}
\end{figure*}

We further investigate the relationship between the disk flux and the inner disk color temperature ($kT_{\rm in}$) during the reflare, as shown in the left panel of Figure~\ref{Fig6. flux_T}. 
We use the Orthogonal Distance Regression (ODR) in the scipy.odr package in result fitting (\url{https://docs.scipy.org/doc/scipy/reference/odr.html}).
The disk flux is obtained from the integral of \textit{diskbb} in $0.1-150$keV.
For fitting these relations, we use the relation Flux $\varpropto$ $
kT_{\rm in}^{\alpha}$ to measure whether the inner radius of the disk has changed during the reflare.
Obviously, we find that it is not possible to describe all observations with one power-law function. 
And then, we try to use a broken power-law model to fit all points.
We find the parameter $\alpha$ deviates far from 4 when the flux above $3.3\pm0.2 \times10^{-9}$ erg cm$^{-2}$ s$^{-1}$ and the disk temperature > 0.23 keV. Below this flux, we find the remaining points follow Flux $\varpropto$ $
kT_{\rm in}^{4.3\pm0.7}$ well.
We also give an unabsorbed X-ray luminosity of $\sim2.5\times10^{36}$ (D/2.2 kpc)$^{2}$ erg s$^{-1}$.
Deviating from this relationship means that the $R_{\rm in}$ and hardening factor change significantly along the flux.
Therefore, using the fitting results of $R_{\rm in}$ and hardening factor $f_{\rm color}=T_{\rm col}/T_{\rm eff}$ from the \textit{kerrd} model, we can study the relation of the disk flux and the $R_{\rm in}^{2}T_{\rm eff}^4$, where $T_{\rm eff}=T_{\rm in}/f_{\rm col}$ is the effective disk temperature. 
In the right panel of Figure~\ref{Fig6. flux_T}, we find all points follow the Flux $\varpropto$ $ (\sqrt{R_{\rm in}}T_{\rm eff})^{4.1\pm 0.2}$, which is consistent with the black body emission from the inner region of the accretion disk.

\subsection{Timing analysis}

\begin{figure*}
    \centering\includegraphics[width=0.49\textwidth]{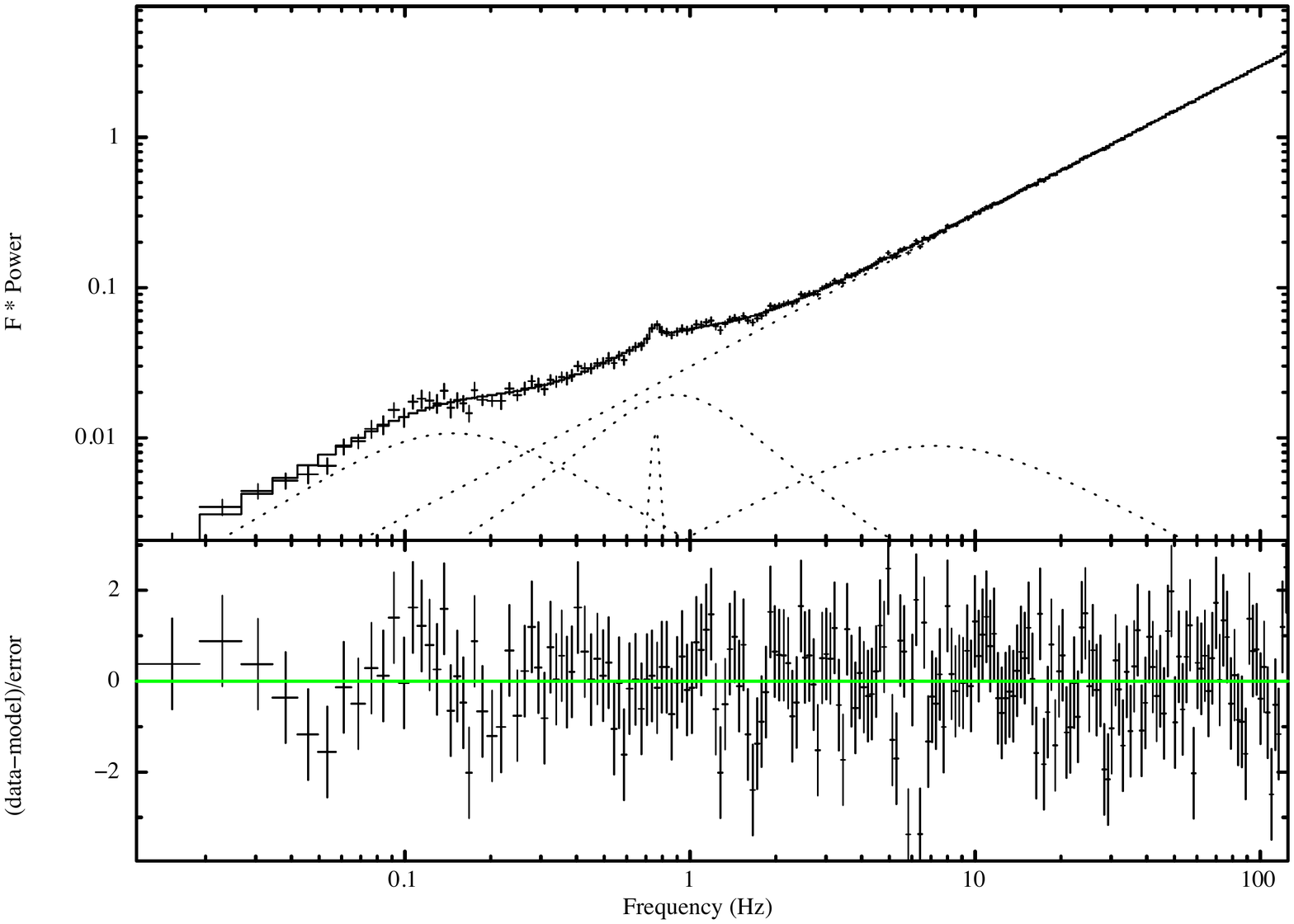}
    \centering\includegraphics[width=0.49\textwidth]{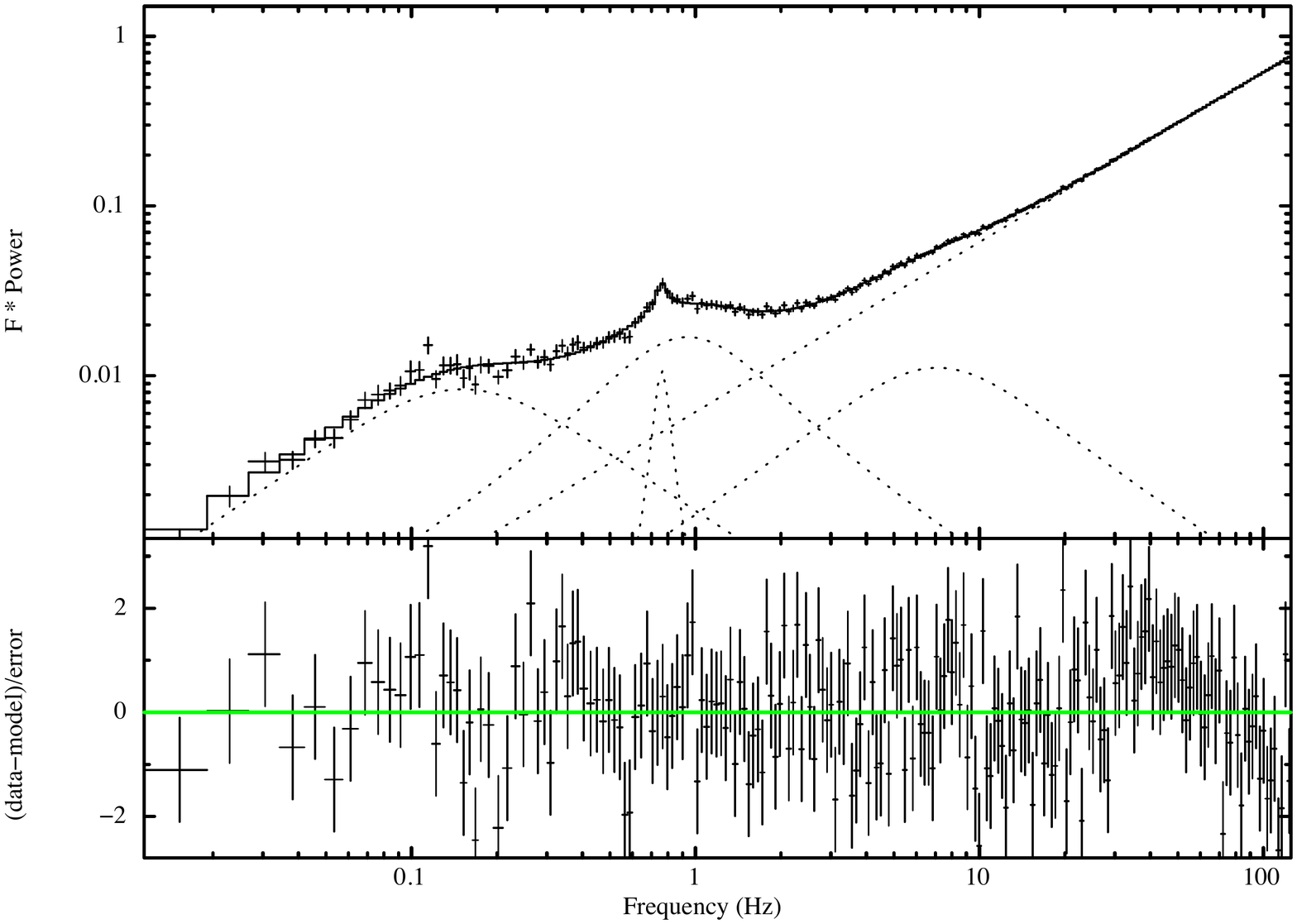}
    \caption{Power density spectra fitting results for ME (right: $10-30$ keV) and HE (left: $25-80$ keV). Here, we just choose a PDS on MJD 58657 with a relatively high confidence level, which shows the noise and QPO components clearly.}
    \label{Fig7. PDS}
\end{figure*}
We study the fast X-ray variability in different energy bands and produce the Power Density Spectrum (PDS) for each observation used for spectral analysis. 
The PDS is applied to Miyamoto normalization. Observations in MJD 58657, 3 days after the peak flux are shown in Figure~\ref{Fig7. PDS}.
It shows the noise and QPO components clearly with a relatively high confidence level. We use multiple Lorentzian components to fit the PDS, where one component can fit a type-C QPO with a quality factor (Q$\equiv$ QPO's central frequency/FWHM) above 2.
We take into account the white noise in the PDS and model it by a power-law component with zero indexes.

\begin{figure*}
    \centering\includegraphics[width=0.49\textwidth]{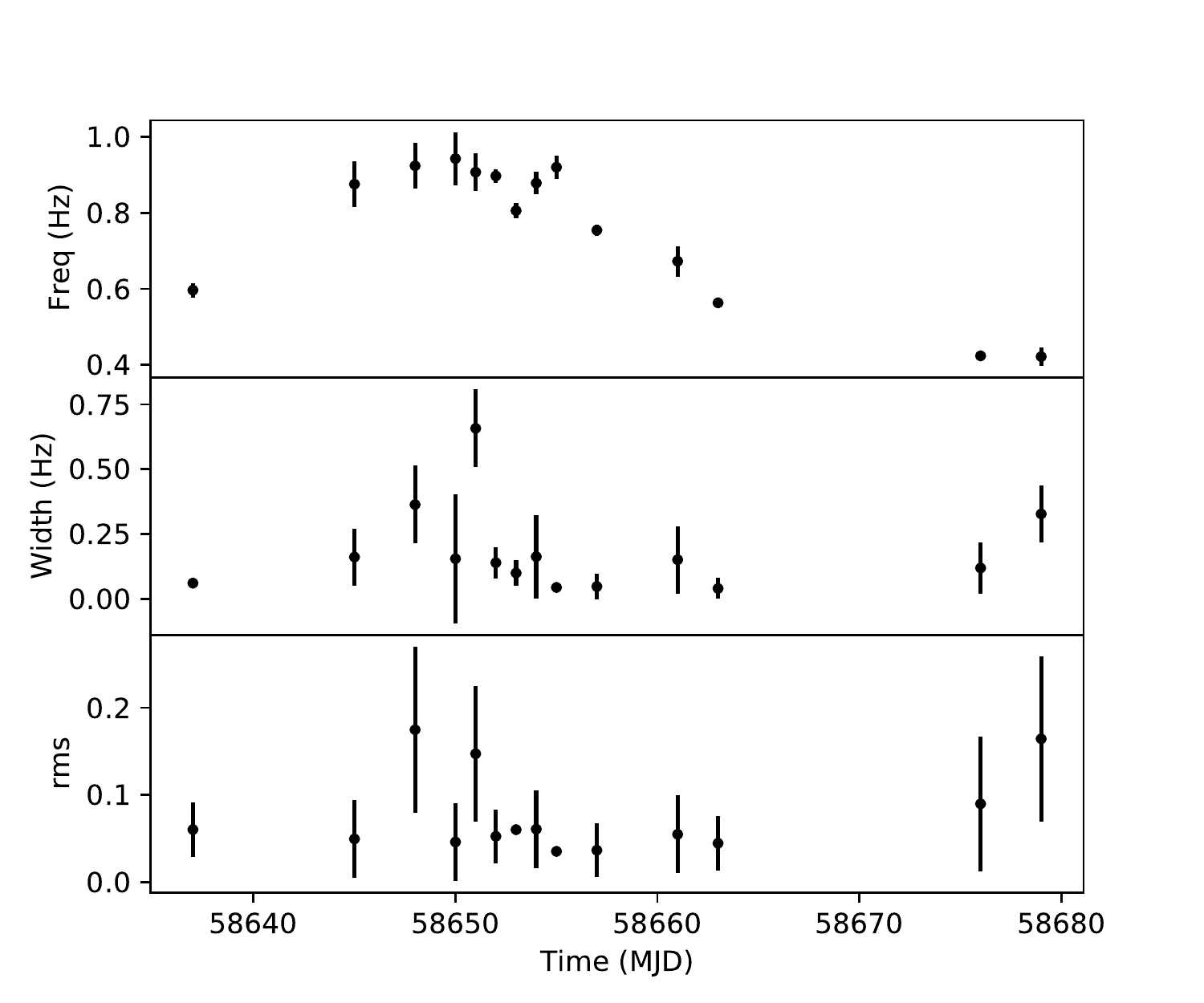}
    \centering\includegraphics[width=0.49\textwidth]{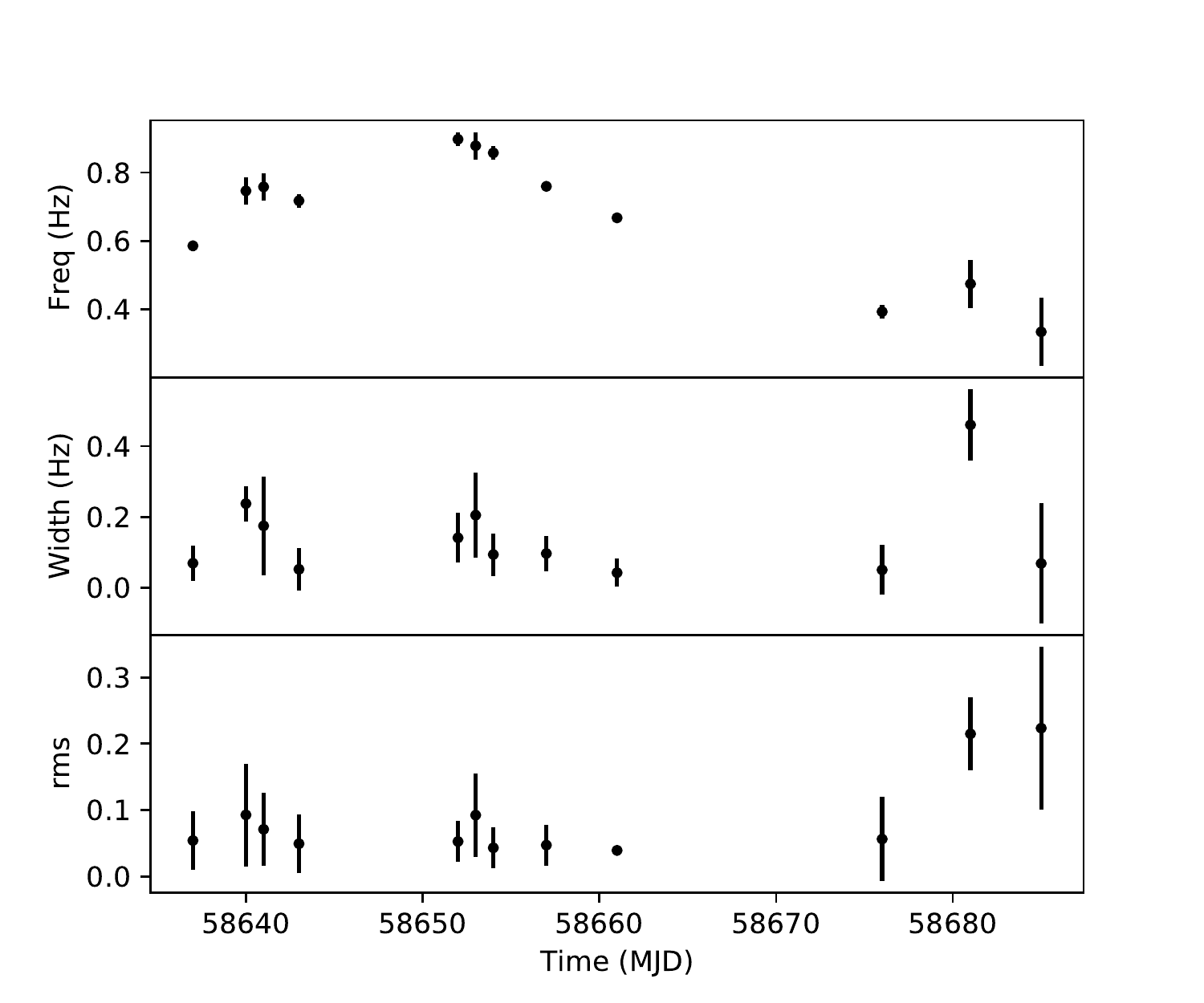}
    \caption{C-QPO frequency, width, and rms as a function of time for ME ($10-30$keV) and HE ($25-80$keV).}
    \label{Fig8. QPO evolution}
\end{figure*}

Based on the data quality, only the combined observations consisting of multiple \textit{Insight}-HXMT exposures provided clearer PDS results.
Meanwhile, only a few observations displayed obvious QPO structures.
We obtained the QPO frequency, and the QPO width, and calculate the QPO fractional RMS throughout the reflare. Figure~\ref{Fig8. QPO evolution} shows the evolution of these three parameters with time in the ME ($10-30$ keV) and HE ($25-80$ keV) energy bands. 
As for \textit{Insight}-HXMT, the ME, and HE energy band offers much more photon counts than the LE energy band, and considering the limited number of QPOs we find at the same time, we ignore the related evolution in the LE energy band. Also, we note that the QPOs in \textit{NICER} were reported in \cite{Alabarta2022MNRAS}.
In Figure~\ref{Fig8. QPO evolution}, although the number of QPOs is limited, the QPO frequency evolution with the flux is evident. 
The type-C QPO has increased from $\sim$ 0.4 Hz, at the beginning and end of the reflare, to 0.9 Hz at the reflare peak.
And the QPO width and the QPO RMS do not vary with flux.
Except for some large error intervals due to low statistical observations, the RMS nearly contains all values within 5\%  in the ME and the HE band, which is lower than RMS $\sim$ $15-20$\% in the $2.5-5.0$ keV energy range in \cite{Alabarta2022MNRAS}. 
Hence, our results show that the QPO RMS decreases to a lower value and then remains stable in the higher energy band.

\begin{figure*}
    \centering\includegraphics[width=0.49\textwidth]{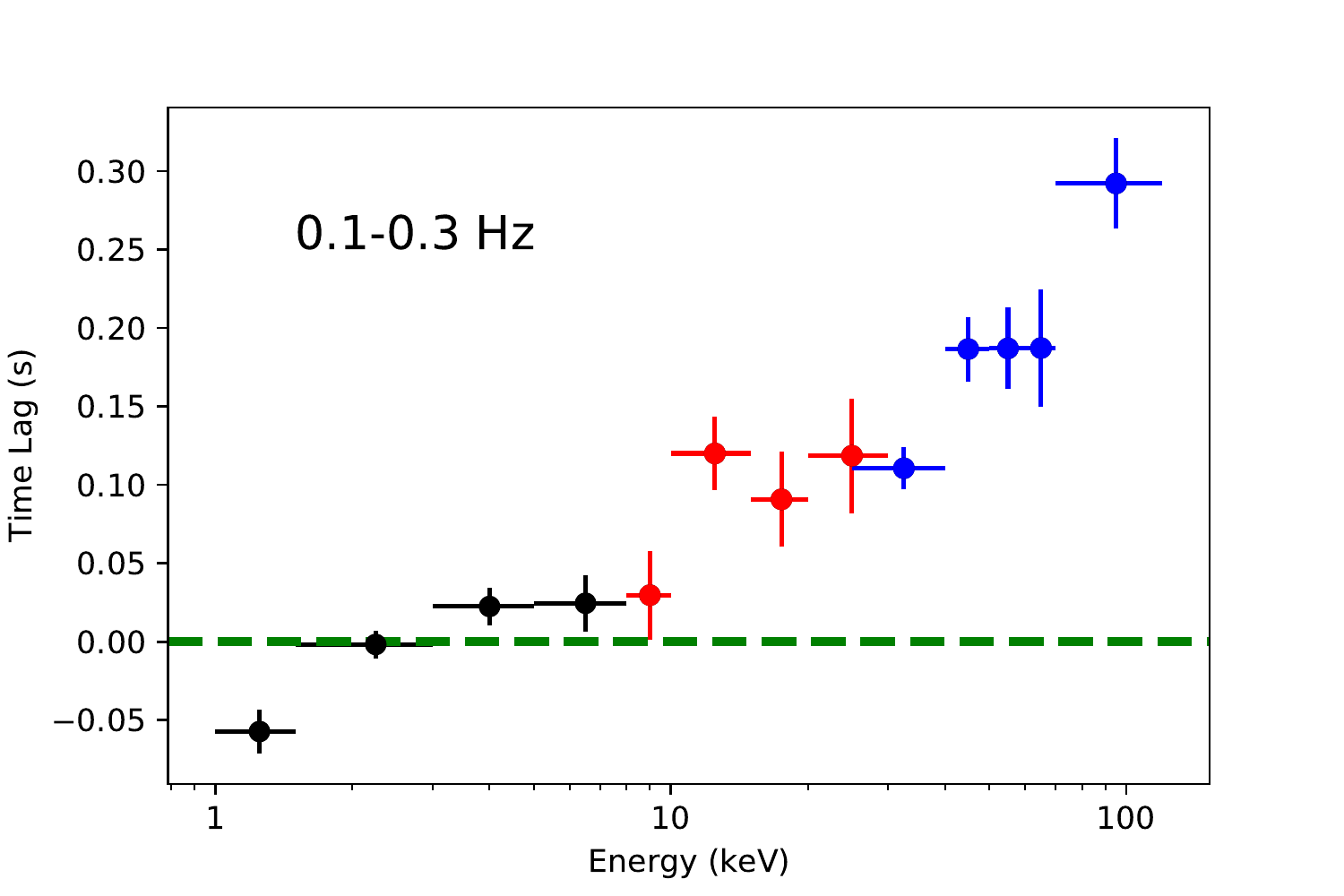}
    \centering\includegraphics[width=0.49\textwidth]{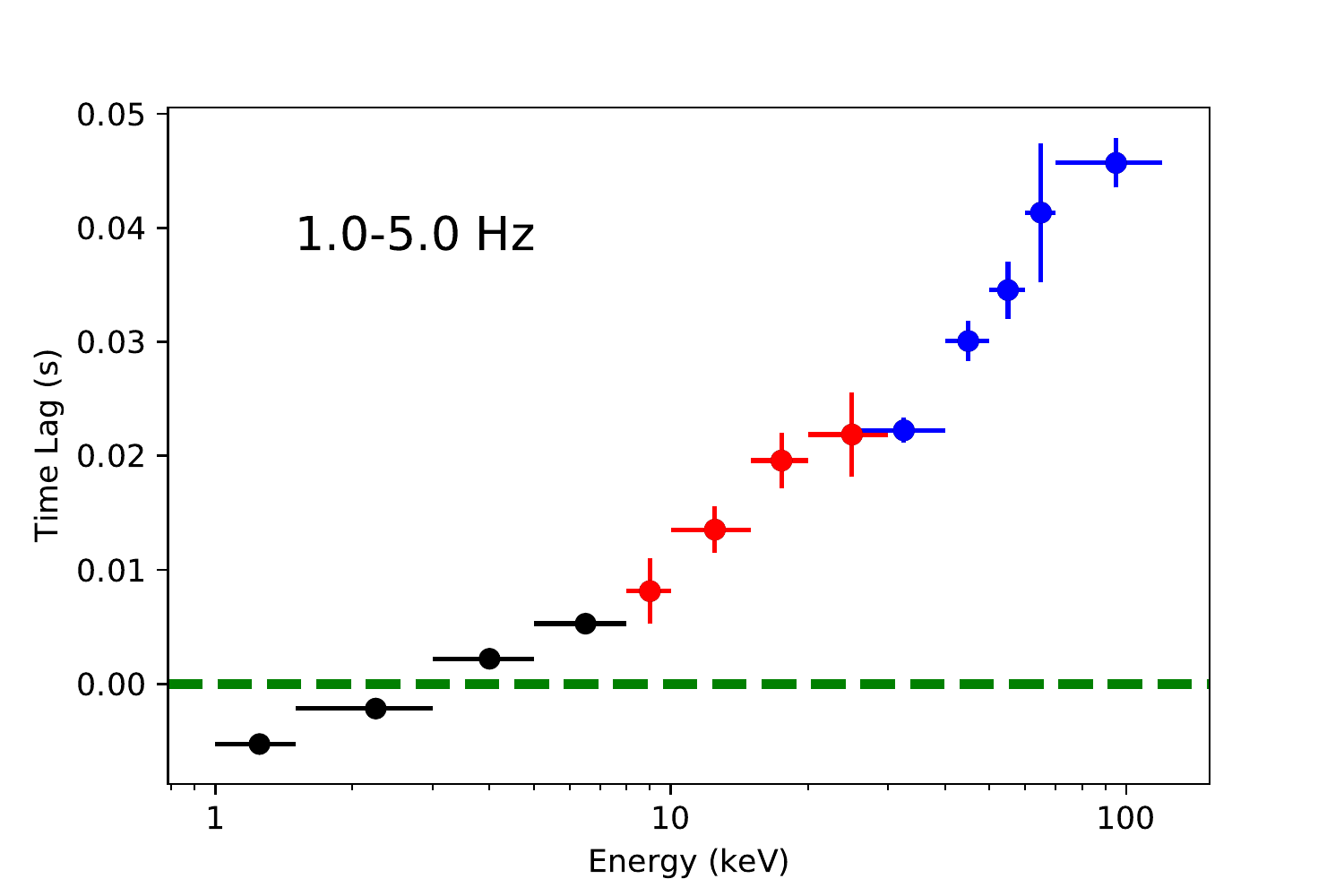}
    \caption{Time lags in different energy bands (reference energy band $1-10$ keV): black for LE ($1.0-1.5$, $1.5-3.0$, $3.0-5.0$, $5.0-8.0$ keV), red for ME ($8.0-10.0$, $10.0-15.0$, $15.0-20.0$, $20.0-30.0$ keV) and blue for HE ($25.0-40.0$, $40.0-50.0$, $50.0-60.0$, $60.0-70.0$, $70.0-120.0$ keV) in the disparate frequency range from HXMT observations on MJD 58657. Left: in $0.1-0.3$ Hz; 
    Right: $1.0-5.0$ Hz. }
    \label{Fig9. Time lag in different energy band}
\end{figure*}

We also study the energy-dependent time lags in two different frequency ranges ($0.1-0.3$ Hz and $1.0-5.0$ Hz respectively, away from the QPO frequency), from 1.0 to 120 keV with respect to the $1.0-10.0$ keV band (reference band). Results are shown in Figure~\ref{Fig9. Time lag in different energy band}. 
The black points, red points, and blue points correspond to LE ($1.0-1.5$, $1.5-3.0$, $3.0-5.0$, $5.0-8.0$ keV), ME ($8.0-10.0$, $10.0-15.0$, $15.0-20.0$, $20.0-30.0$ keV) and HE ($25.0-40.0$, $40.0-50.0$, $50.0-60.0$, $60.0-70.0$, $70.0-120.0$ keV), respectively (see again Figure~\ref{Fig9. Time lag in different energy band}). 
A hard lag (hard photons delay soft photons) occurs, and the time lags increase with the energy band.
In $0.1-0.3$ Hz, the time lag in the high energy band between $70-120$ keV is $\sim\ 0.29$ s, and the time lag in $1.0-5.0$ Hz between $70-120$ keV is $\sim\ 0.05$ s.

\begin{figure*}
    \centering\includegraphics[width=0.49\textwidth]{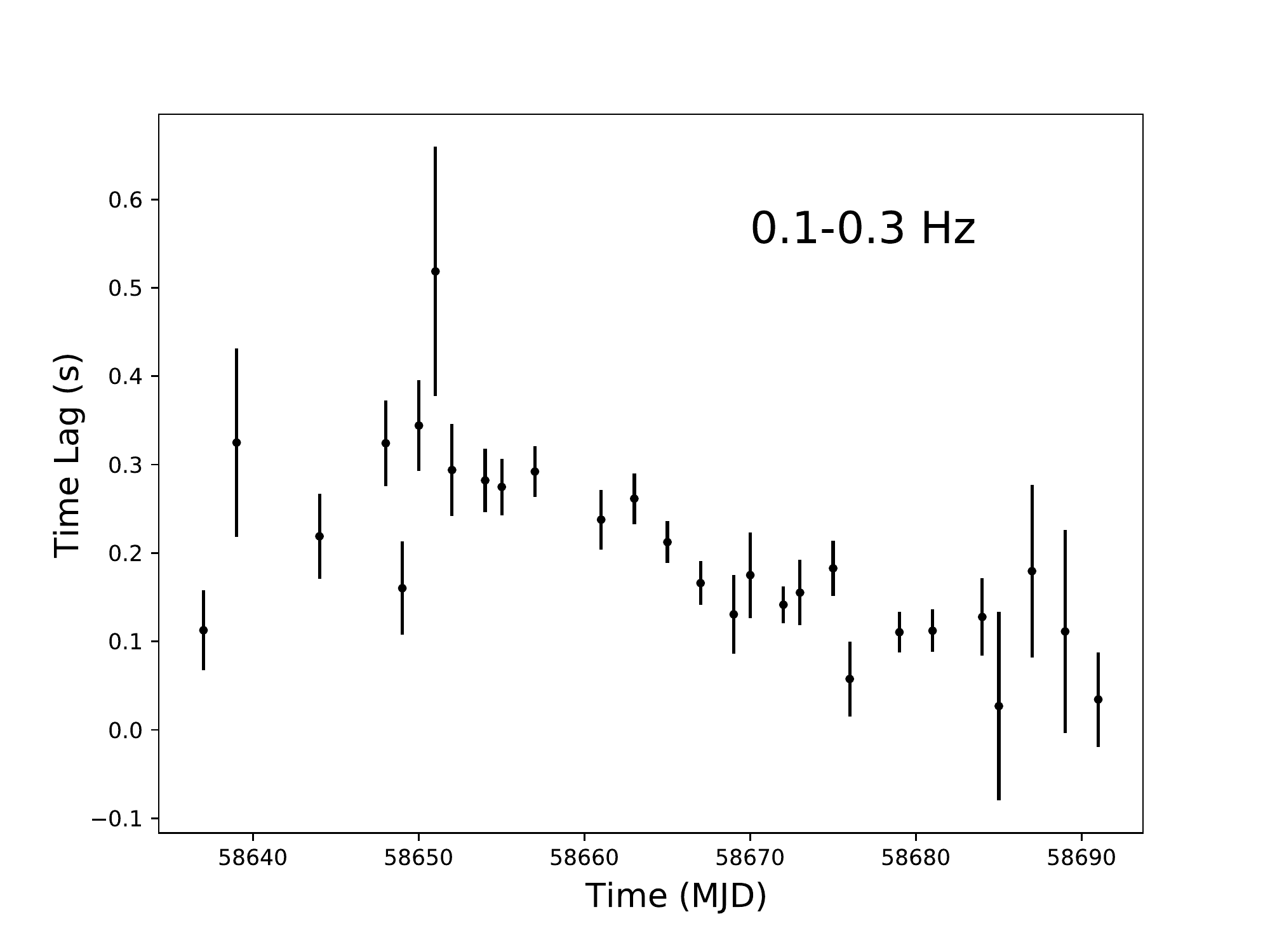}
    \centering\includegraphics[width=0.49\textwidth]{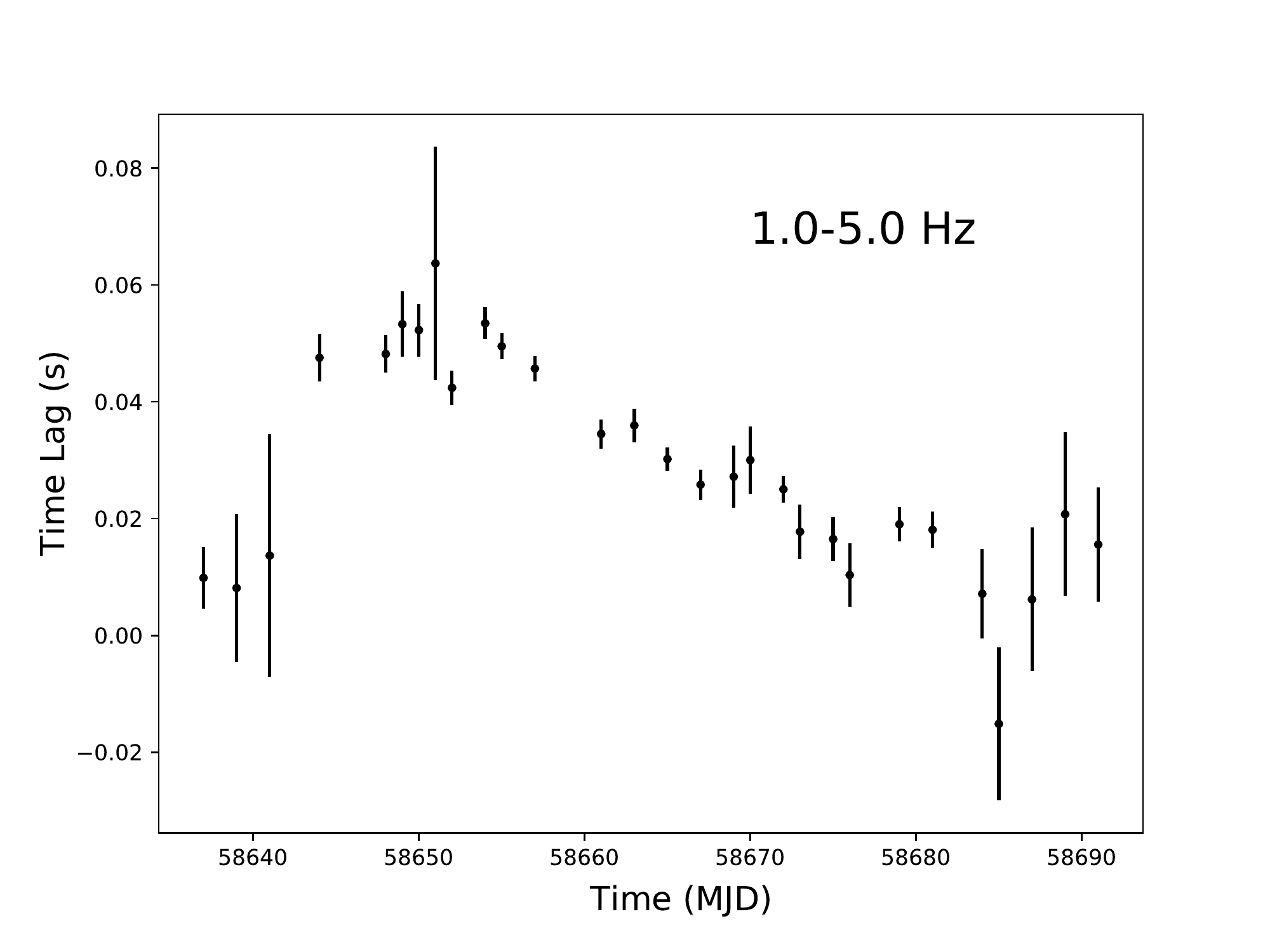}
    \caption{Time lag evolution in high energy band $70.0-120.0$ keV for two frequency ranges. Left: in $0.1-0.3$ Hz; 
    Right: $1.0-5.0$ Hz..}
    \label{Fig10. Time lag evolution in 70-120 keV}
\end{figure*}

Time lags are also time dependent. The evolution of time lags in $70-120$ keV with the largest time lag, are shown in Figure~\ref{Fig10. Time lag evolution in 70-120 keV}. 
When frequency is lower or higher than QPO frequency (i.e. $0.1-0.3$ Hz and $1.0-5.0$ Hz), the evolution of time lags is not affected by QPO components. As seen in left panel and right panel of Figure~\ref{Fig10. Time lag evolution in 70-120 keV} , in $0.1-0.3$ Hz, along with increasing flux, the time lags increase from 0.1 s to 0.4 s.
Also, compared to low frequency, we find that the time lags at $1.0-5.0$ Hz evolve from 0 s to 0.06 s. Here, we notice that although different frequency ranges are selected, the phase delays are similar. So this suggests that the physical reason for this delay at different time scales could be same.

\begin{figure*}
    \centering\includegraphics[width=0.49\textwidth]{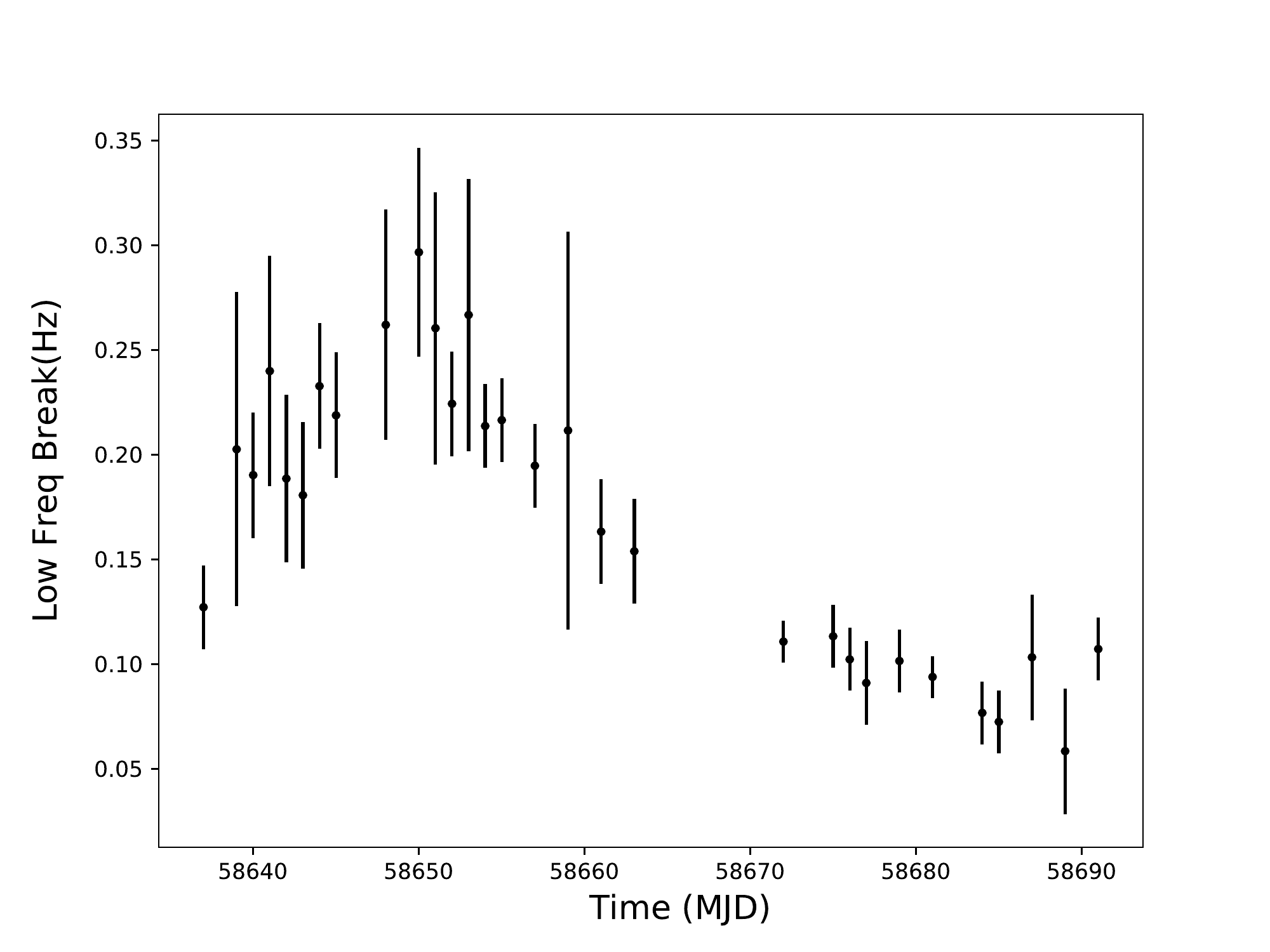}
    \centering\includegraphics[width=0.49\textwidth]{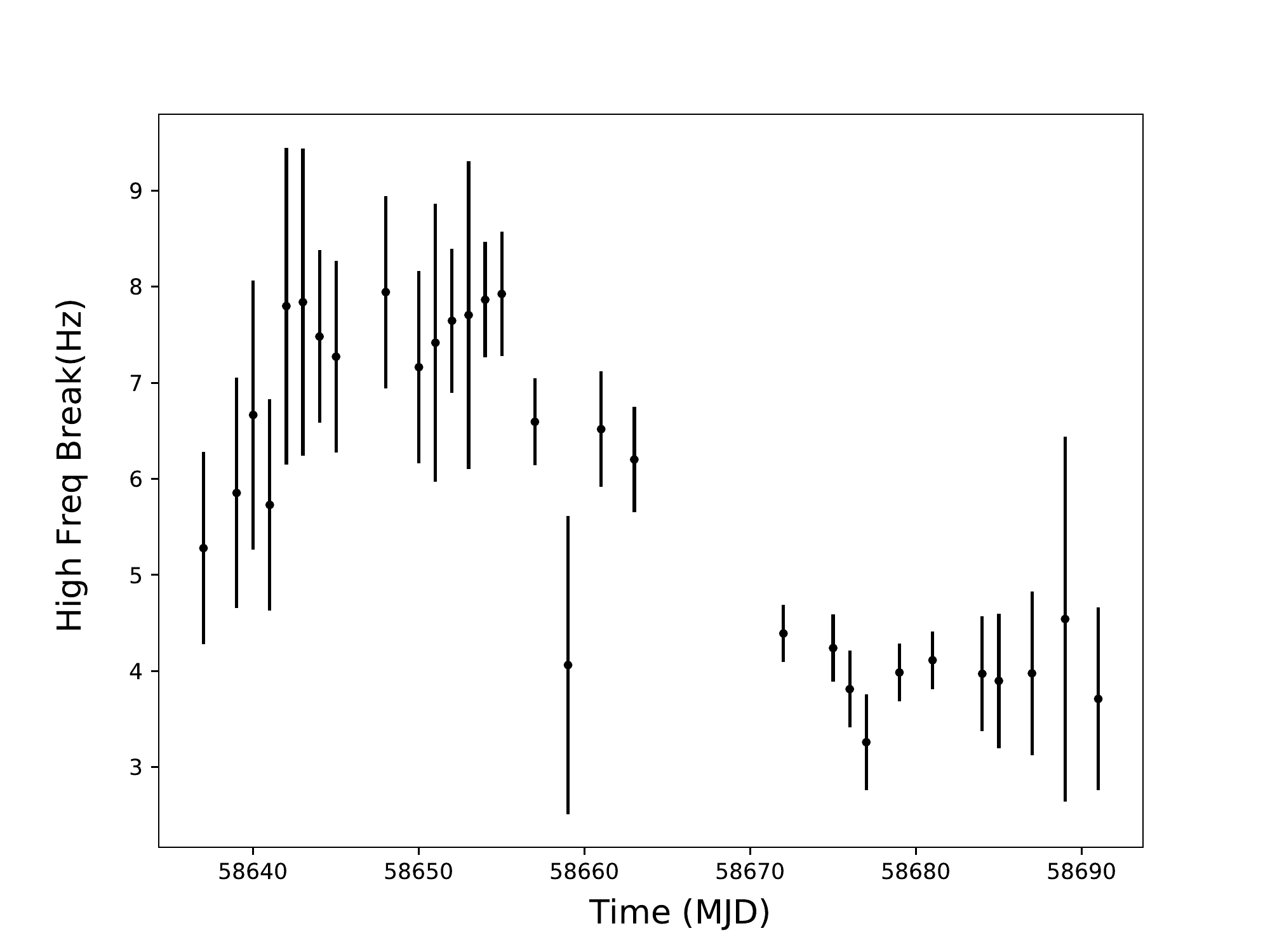}
    \caption{
    Left and right panels present the evolution of the low frequency break in high energy and high frequency break in high energy, respectively.}
    \label{Fig11. Cutoff frequency evolution}
\end{figure*}

For further analysis, in Figure~\ref{Fig11. Cutoff frequency evolution}, we show the evolution of the break frequency or characteristic frequency for the different noise components in Figure~\ref{Fig7. PDS}.
The left and right panel show the lowest characteristic frequency (low frequency break) and highest characteristic frequency (high frequency break) in the high energy band ($25-80$ keV), respectively.
We find that the high-energy band shows pronounced evolutionary trends, i.e., the characteristic frequency increases as the flux increases, and we also find that the characteristic frequency of the high-frequency noise has the same evolutionary trend as the low-frequency noise.

\section{Discussion}

\subsection{Spectrum evolution}
In this paper, we focus on the first reflare after the main outburst of the low mass black hole X-ray binary (LM-BHXRB) MAXI J1348-630 in 2019, and in particular on the spectral and timing evolution throughout the reflare. This is the first time that the reflare properties are studied in a broad energy band ($0.3-150$ keV) by joint \textit{NICER} and \textit{Insight}-HXMT observations. 

In all our observations, thermal emission from an accretion disk is required by the fitting residual observed using a non-thermal model only.
When the non-thermal component dominates, the scattering of soft photons in the corona dominates.
To avoid underestimating the soft photons from the disk, we use the convolution model \textit{thcomp} to fit the non-thermal components and their internal scattering processes. 
In Figure~\ref{Fig3. Parameters_diskbb}, the column density $n_{\rm H}$ stays at $\sim\ 0.86\times10^{22}$ cm$^{-2}$ during the reflare, except for some observations around the peak. Considering that the residual below $2.1$ keV from \textit{NICER}, can be affected by the high flux changing apparently the absorption parameter $n_{\rm H}$,  we fix its value to that of $n_{\rm H}$ during the soft state (SS) estimated by Swift/XRT (\citealp{Tominaga2020ApJ}), and the average $n_{\rm H}$ during the whole main outburst (\citealp{ZhangW2022ApJ}). 
We note that the $N_{\rm disk}$ is correlated with the $kT_{\rm in}$, while the cover factor $Cov_{\rm f}$ shows slight anti-correlation with flux.
The $N_{\rm disk}$ trend indicates that the inner radius of the accretion disk recedes as the accretion rate increases, which is clearly inconsistent with the expected evolution of the canonical accretion disk: the inner radius should become smaller, or keep stable at the ISCO when the accretion rate increases.
We note that such a peculiar trend was also reported in \cite{Zhang2020MNRAS}, who used only \textit{NICER} observations. 
Through the joint analysis of \textit{NICER} and \textit{Insight}-HXMT, we find that this trend is not due to the narrow energy band used for the fitting.
Meanwhile, in \cite{ZhangW2022ApJ}, an outward trend of the inner radius was also found during the hard state (HS) of the main outburst of MAXI J1348-630, even using self-consistent Comptonization models, which take into account the upscattering of photons from the disk emission into the corona region. 
They suggested that at the beginning of the outburst, the hardening factor for the inner-disk emission could be larger than the canonical value $\sim$1.7 of the standard disk (\citealp{Shimura1993ApJ, Shafee2006ApJ}).  
\cite{Dunn2011MNRAS} found that the hardening factors can reach a value as high as $\sim$3 in the initial phase of an outburst, which can be used to explain the anomalously low values of $R_{\rm in}$ in some black hole transients.
We suggest that the same process occurs in the reflare; the detected trend of increasing $N_{\rm disk}$ is caused by the hardening factor deviating far from 1.7 from a sufficiently high effective optical thickness of the standard Shakura–Sunyaev disk (\citealp{Shakura1973AA}).
We note that MAXI J1348-630 is a Kerr black hole because of its high spin parameter $a=0.8-0.998$ (\citealp{Kumar2022MNRAS, Jia2022MNRAS}). So, we replace the \textit{diskbb} model with \textit{kerrd} model which describes the thermal emission from the optically thick accretion disk around a Kerr black hole. The model incorporates the spectral hardening factor, $T_{\rm col}/T_{\rm eff}$ and $R_{\rm in}$ for a given distance (fixed at 2.2 kpc), inclination angle (fixed at $36.5^\circ$)and black hole mass (fixed at 8.7$M_{\rm \odot}$). 
In Figure~\ref{Fig4. Parameters_kerrd}, if we fix the $T_{\rm col}/T_{\rm eff}$ at 1.7, we get the $R_{\rm in}$ increase as the accretion rate becomes larger, which is similar to the trend of $N_{\rm disk}$ in Figure~\ref{Fig3. Parameters_diskbb}.
If we free $T_{\rm col}/T_{\rm eff}$, the evolutionary trend of $R_{\rm in}$ changes significantly.
The $R_{\rm in}$ decreases from 34 $R_{\rm g}$ to 17 $R_{\rm g}$ with the increasing accretion rate within MJD $58637-58673$, while at a lower accretion rate the increased accretion rate does not lead to a significant change of $R_{\rm in}$ and it stays at 34 $R_{\rm g}$.
The $R_{\rm in}$ estimated from the \textit{kerrd} model is similar to the $R_{\rm in}$ estimated from the reflection model (\citealp{Kumar2022MNRAS}).
As discussed earlier, the change in $R_{\rm in}$ is due to the evolution of the $T_{\rm col}/T_{\rm eff}$ from $\sim$3 to $\sim$1.7, and similar to the $R_{\rm in}$, there is a clear decline for the $T_{\rm col}/T_{\rm eff}$ only in MJD $58640-59673$, while it is stable at $\sim$3 at other times.

In Figure~\ref{Fig6. flux_T}, we plot the variation in the disc flux with the disc temperature $kT_{\rm in}$, and we also plot the disc flux with $\sqrt{R_{\rm in}}kT_{\rm eff}$.
In the left panel, We find that the overall evolutionary trend is not consistent with the Flux $\propto kT_{\rm in}^4$. 
However, the points below the red dashed line can be fitted by Flux $\propto kT_{\rm in}^{4.3\pm0.7}$.
For observations with flux above $3.3\pm0.2\times10^{-9}$ erg cm$^{-2}$ s$^{-1}$, as disk flux increases, the $kT_{\rm in}$ stays nearly stable at 0.23 keV.
Based on these results, we find that for observations with low fluxes (flux below the red dashed line in Figure~\ref{Fig6. flux_T}), the disk does not change significantly. 
In the right panel, we consider the effect of the hardening factor (color correction factor) $f_{\rm color}=T_{\rm col}/T_{\rm eff}$ on disk temperature and the evolution of the inner radius $R_{\rm in}$. So, it shows the true physical temperature versus flux evolution of the accretion disk, following the Flux $\propto R_{\rm in}^2T_{\rm eff}^{4}$.
In Figure~\ref{Fig4. Parameters_kerrd}, we note that the purple points in the left panel of Figure~\ref{Fig6. flux_T} correspond to observations where the hardening factor is stable at $\sim3$ and the inner radius is stable at 34 $R_{\rm g}$. 

From the spectral analysis, we find that the evolution of the inner radius of the accretion disk with the accretion rate is not a continuously decreasing process, but can be divided into two different regions according to a critical luminosity $L_{\rm crit}$.
In the region below the $L_{\rm crit}$, the inner radius of the disk does not change significantly and is truncated at a relatively far distance from the black hole. 
In the region above the $L_{\rm crit}$, on the other hand, the disk evolvs with the usual expected HS, with the inner radius gradually approaching the $R_{\rm ISCO}$ as the accretion rate increases. 
The studied reflare only goes through a HS. Therefore, although the inner disk moves inward, it does not reach the ISCO.
From the luminosity of the source at the red dashed line in Figure~\ref{Fig6. flux_T}, this characteristic luminosity can be obtained as $L_{\rm crit}\sim2.5\times10^{36}$ erg s$^{-1}\approx0.002\ L_{\rm Edd}$ (calculated in 1-10 keV), and the $L_{\rm Edd}=1.3\times10^{38}(M/M_{\odot})$ erg s$^{-1}$, where $M=8.7\ M_{\odot}$. 
In \cite{Shakura1973AA}, from the standard disk solution, the hardening factor $f\sim(\alpha^{es}/\alpha^{ff})^{1/8}\propto n_{\rm e}^{-1/8}$, where $\alpha^{es}$ is the electron-scattering coefficient, $\alpha^{ff}$ is the free–free absorption coefficient and $n_{\rm e}$ is the electron density.
During the reflare, except for the peak with $f\sim1.7$ which is the same as the characteristic value of a standard disk, in most observations the inner region of the accretion disk might still be in a non-equilibrium state, that is, it is still in the process of condensing from the hot corona (\citealp{LiuBF2007ApJ, LiuBF2011ApJ, Taam2008ApJ, Meyer-Hofmeister2009AA, Qiao2017MNRAS}). 
Combined with Figure~\ref{Fig4. Parameters_kerrd} and Figure~\ref{Fig6. flux_T}, we speculate that, for the luminosity above the $L_{\rm crit}$, the cooling of the corona becomes more dominant, and the inner disc gradually becomes a standard disk; for the luminosity below the $L_{\rm crit}$, the evaporation process of the disk might be more dominant, resulting in the inability of the accretion material in the inner region of the disk to form an effective inner flow of material and transfer the angular momentum outward.
The hardening factor f can also be determined by the Comptonization in the corona near or covering the accretion disk (\citealp{Shimura1993ApJ}), and such process can be described by the cover factor $Cov_{\rm f}$ in the \textit{thcomp} model. 

In Figure~\ref{Fig4. Parameters_kerrd}, at the end of the reflare, we find that the $Covf_{\rm f}$ recovers simultaneously with the f. This suggests that there is indeed a connection between the cover factor and hardening factor, but $Cov_{\rm f}$ is clearly not the only reason affecting the f, since $Cov_{\rm f}$ begins at a lower value, suggesting that f is also influenced by other factors.

\subsection{QPO evolution}
In Figure~\ref{Fig8. QPO evolution}, we firstly show the type-C QPO evolution in the higher energy band during a reflare. The frequencies of the QPO are consistent with the \textit{NICER} observations (\citealp{Alabarta2022MNRAS}). Compared to their results (Figure 9 in \cite{Alabarta2022MNRAS}), our results show that the QPO rms keeps at $\sim5\%$, and it does not evolve significantly with energy and flux above 10 keV. 
Combined with the \textit{NICER} results in \cite{Alabarta2022MNRAS}, we find that the shape of the QPO rms spectrum evolves during the reflare. 
In \cite{Alabarta2022MNRAS}, for higher flux observations, the RMS spectrum increased with energy below 2.5 keV and stayed at the maximum value $\sim10\%-15\%$ between $2.5-7$ keV. Then we find it decreases to $5\%$ above 7 keV.
For lower flux observations, the RMS spectrum stabilized at $\sim20\%$ below 7 keV and decreased sharply to $5\%$ above 7 keV. 
The trends above 2.5 keV are similar to the results of the failed-transition outburst of H 1743-322 (\citealp{Wang2022MNRAS}) and the first outburst of MAXI J1820+070 (\citealp{Ma2021NatAs}).
These findings suggest that the trend is typical for sources that have been in a hard state for a long time during an outburst. 
While the trends below the 2.5 keV were similar to the trends below 10 keV of the intermediate state of the canonical black hole outburst. 

For the shape of QPO's RMS with energy, we speculate that it is determined by two aspects: on the one hand, it is related to the flux ratio of the spectrum components contributing RMS to the total energy spectrum at different energy bands, and on the other hand, it is determined by the intrinsic properties of QPO itself (\citealp{Gierlinski2005MNRAS, Sobolewska2006MNRAS, Kong2020JHEAp, Shui2021MNRAS}). 
If the QPO originates from the Lense-Thirring (L-T) precession of the corona, disk does not contribute to the QPO RMS (\citealp{Kong2020JHEAp}). Based on the above description, we can interpret the QPO RMS spectrum. 
Firstly, for energies above 2.5 keV, the flux ratio of the non-thermal component to the total energy spectrum does not change with energy because the flux of the disk is weak, and the geometric properties of the corona determine the RMS size. For non-thermal radiation, we assume that the high-energy photons arise from the inner region of the corona, while the low-energy photons arise from the outer region of the corona. If the geometry of the corona is considered to be  more similar to the flat shape of a disk in the outer region, and a more spherical shape in the inner region, then the projection effect is larger in the outer region, which would naturally explain the observed QPO rms spectra.
We also note that the higher energy band of RMS spectra could also originate from the small-scale jets procession (\citealp{Ma2021NatAs}), so a low-speed stable jet could also explain the high-energy part of the RMS energy spectrum.
Secondly, for energies below 2.5 keV, the QPO rms decreases with higher flux. Because the inner radius of the inward accretion disk can contribute to a higher disk flux and disk temperature, the flux ratio of the non-thermal component to total spectral will decrease in the low energy band, hence lowering the QPO rms. 
For low flux, the far truncated non-standard disk dominated by the evaporation allowed the QPO rms to be dominated by the corona in all energy bands.

\subsection{Noise evolution}

Further analyses of the noise time lag between high energy photons ($70-120$ keV) with reference to low energy photons ($1-10$ keV) in the low and high frequency range are shown in Figure~\ref{Fig10. Time lag evolution in 70-120 keV}. 
For low and high-frequency ranges, the hard lag evolution shows significant evolution between MJD 58640 and MJD 58670. We find that this time range corresponds exactly to period of $R_{in}$ evolution in Figure~\ref{Fig4. Parameters_kerrd}, when the inner radius of the disk moves significantly inward. For the rest of the time, the hard lag remains stable.
Therefore, the hard lag reflects the Compton upscattering process of low energy photons in the corona. 
For low flux range, the small delay might indicate that both low-energy and high-energy photons experience sufficient Compton scattering before they leave the corona, which is consistent with the high coverage in Figure~\ref{Fig4. Parameters_kerrd}. 
Also, if the disk is considered to deviate from the standard disk in low flux, the small delay indicates that photons radiating from inside the disk or corona are difficult to distinguish.
In contrast, for the high flux range, a low coverage corona with a cooling/condensing-dominated standard optical thick disk will be able to effectively distinguish between photons from the disk or the corona, producing a significant hard delay. At this point, the larger hard delay with increasing flux may indicate that more scattering time is required for the photons to scatter from low energy (1-10 keV) to high energy (70-120 keV), which may be related to the effective cooling of the corona by the disk. This is consistent with the results in Figure~\ref{Fig4. Parameters_kerrd} where the electron temperature is lower at high accretion rates.

The low and high characteristic frequencies of two noise components are shown in Figure~\ref{Fig11. Cutoff frequency evolution}. 
We find that, similar to the time lag in Figure~\ref{Fig10. Time lag evolution in 70-120 keV}, the characteristic frequencies can also be divided into two different regions because the characteristic low-frequency break can be associated with the viscous time-scale ($t_{\rm visc}$) at the truncation radius of the accretion disc (\citealp{Done2007AARv, Husain2022MNRAS}). 
Meanwhile, the $t_{\rm visc}\propto R_{\rm in}^{3/2}$, so the constant viscous timescale means that $R_{\rm in}$ does not change. 
The characteristic high-frequency break might be associated with the dynamical time scale ($t_{\phi}$) at the truncation radius. Because of $t_{\phi}\sim \alpha (H/R)^2 t_{\rm visc}$, in Figure~\ref{Fig11. Cutoff frequency evolution} the right panel also shows a similar trend to the left panel.

\section{Conclusions}

In this paper, we discussed the spectral and timing characteristics of the LM-BHXRB MAXI J1348-630 during its second outburst, or reflare. 
Our results support the existence of a characteristic luminosity, $L_{\rm crit}\sim0.002\ L_{\rm Edd}$ that makes the inner radius of the disk exhibit different properties as the accretion rate increases. 
The inner radius of the disk far from the black hole is kept constant below the $L_{\rm crit}$, which might be related to the dominant evaporation process in the accretion disk. When the condensing process dominates above the $L_{\rm crit}$, the disk formation will be very effective, and angular momentum transfer can occur. 
Further confirmation for this scenario requires future X-ray telescopes with large areas and high sensitivity at both low energy and high energy bands simultaneously, such as eXTP.

\section*{Acknowledgements}

This work is supported by the National Key R\&D Program of China (2021YFA0718500), the National Natural Science Foundation of China under grants U1838201, and Guangdong Major Project of Basic and Applied Basic Research (Grant No. 2019B030302001).
This work was partially supported by International Partnership Program of Chinese Academy of Sciences (Grant No.113111KYSB20190020). 
This work is supported by China Scholarship Council (CSC) at Eberhard Karls Universit{\"a}t of T{\"u}bingen.

\section*{Data Availability}

This work made use of data from the \textit{Insight-HXMT} mission (\href{http://hxmtweb.ihep.ac.cn/}{http://hxmtweb.ihep.ac.cn/}), a project funded by China National Space Administration (CNSA) and the Chinese Academy of Sciences (CAS). The data of \textit{NICER} obtained through the High Energy Astrophysics Science Archive Research Center (HEASARC, \href{https://heasarc.gsfc.nasa.gov/}{https://heasarc.gsfc.nasa.gov/}).



\bibliographystyle{mnras}
\bibliography{ref} 








\bsp	
\label{lastpage}
\end{document}